%% file: main.tex
\documentclass[fleqn,usenatbib]{mnras}
\pdfoutput=1
\usepackage{newtxtext,newtxmath}
\usepackage{xcolor}
\usepackage{multirow}
\usepackage{threeparttable}
\usepackage{subcaption}
\usepackage{mwe}

\usepackage[T1]{fontenc}

\DeclareRobustCommand{\VAN}[3]{#2}
\let\VANthebibliography\thebibliography
\def\thebibliography{\DeclareRobustCommand{\VAN}[3]{##3}\VANthebibliography}

\usepackage{graphicx,caption}	% Including figure files
\usepackage{amsmath}	% Advanced maths commands

\usepackage{footnote}
\makesavenoteenv{table}
\makesavenoteenv{tabular}

\title[Subpulse Drifting in PSR J1822$-$2256]{Revisiting the sub-pulse drifting phenomenon in PSR J1822$-$2256: Drift Modes, Sparks, and Emission Heights}

\author[P. Janagal et al.]{Parul Janagal,$^{1}$\thanks{E-mail: phd1801121004@iiti.ac.in}
Manoneeta Chakraborty,$^{1}$
N. D. Ramesh Bhat,$^{2}$
Bhaswati Bhattacharya,$^{3}$
\newauthor Samuel J. McSweeney$^{2}$
\\
$^{1}$Department of Astronomy, Astrophysics, and Space Engineering, Indian Institute of Technology Indore, Indore 453552, India\\
$^{2}$International Centre for Radio Astronomy Research, Curtin University, Bentley, WA 6102, Australia\\
$^{3}$National Centre for Radio Astrophysics, Tata Institute of Fundamental Research, Pune University, Pune 411007, India\\
}

\date{Accepted XXX. Received YYY; in original form ZZZ}

\pubyear{2021}

\begin{document}
\label{firstpage}
\pagerange{\pageref{firstpage}--\pageref{lastpage}}
\maketitle

\input{abstract.tex}


%%%%%%%%%%%%%%%%%%%%%%%%%%%%%%%%%%%%%%%%%%%%%%%%%%

%%%%%%%%%%%%%%%%% BODY OF PAPER %%%%%%%%%%%%%%%%%%

\input{introduction.tex}
\input{observations.tex}

\input{analysis.tex}
\input{discussion.tex}
\input{conclusion.tex}

\section*{Acknowledgements}

PJ acknowledges the Junior Research Fellowship awarded by the Council of Scientific \& Industrial Research, India. MC thanks the INSPIRE research grant (DST/INSPIRE/04/2016/001187) awarded under the Department of Science \& Technology, India. We would also like to thank our referee, Geoff Wright, for the detailed comments and suggestions that improved the presentation of the paper. We thank the staff of the GMRT who have made these observations possible. The GMRT is run by the National Centre for Radio Astrophysics of the Tata Institute of Fundamental Research. The Parkes radio telescope is part of the Australia Telescope National Facility which is funded by the Australian Government for operation as a National Facility managed by CSIRO.

%%%%%%%%%%%%%%%%%%%%%%%%%%%%%%%%%%%%%%%%%%%%%%%%%%
\section*{Data Availability}

This paper includes data taken from the uGMRT in the 38th observing cycle. We have also used 1400 MHz data from the Parkes radio telescope available at the public data archive maintained by the CSIRO. The datafile used is available at \url{https://academic.oup.com/mnras/article/474/4/4629/4705909#supplementary-data}
 
%The inclusion of a Data Availability Statement is a requirement for articles published in MNRAS. Data Availability Statements provide a standardised format for readers to understand the availability of data underlying the research results described in the article. The statement may refer to original data generated in the course of the study or to third-party data analysed in the article. The statement should describe and provide means of access, where possible, by linking to the data or providing the required accession numbers for the relevant databases or DOIs.



\bibliographystyle{mnras}
\bibliography{references} % if your bibtex file is called example.bib

%\begin{thebibliography}{99}
%\bibitem[\protect\citeauthoryear{Author} {2012}]{Author2012}
%\end{thebibliography}

%%%%%%%%%%%%%%%%%%%%%%%%%%%%%%%%%%%%%%%%%%%%%%%%%%

%%%%%%%%%%%%%%%%% APPENDICES %%%%%%%%%%%%%%%%%%%%%

%\appendix


%If you want to present additional material which would interrupt the flow of the main paper, it can be placed in an Appendix which appears after the list of references.

%%%%%%%%%%%%%%%%%%%%%%%%%%%%%%%%%%%%%%%%%%%%%%%%%%

% Don't change these lines
\bsp	% typesetting comment
\label{lastpage}
\end{document}

%% file: abstract.tex
\begin{abstract}
Sub-pulse drifting in pulsar radio emission is considered to be one of the most promising phenomenon for uncovering the underlying physical processes. Here we present a detailed study of such a phenomenon in observations of PSR J1822$-$2256,  made using the upgraded Giant Meterwave Radio Telescope (uGMRT). Observations were made simultaneously using the Band 3 (300-500 MHz) and Band 4 (550-750 MHz) receivers of the uGMRT. The pulsar is known to exhibit subpulse drifting, mode changing, and nulling. Our observations reveal four distinct sub-pulse drifting modes of emission (A, B, C, and D) for this pulsar, with the drift periodicities of 17.9 $P_1$, 5.8 $P_1$, 8 $P_1$, 14.1 $P_1$, respectively (where $P_1$ is the pulsar rotation period), two of which exhibit some new features that were not reported in the previous studies. We also investigate the possible spark configuration, characterised by the number of sparks ($n$) in the carousel patterns of these four drift modes, and our analysis suggests two representative solutions for the number of sparks for a carousel rotation period, $P_4$, which lies in the range of $13$ to $16$. The large frequency coverage of our data (300-750 MHz) is also leveraged to explore the frequency dependence of single-pulse characteristics of the pulsar emission, particularly the frequency-dependent subpulse behaviour and the emission heights for the observed drift modes. Our analysis suggests a clear modal dependence of inferred emission heights. We discuss the implications for the pulsar emission mechanism and its relation to the proposed spark configuration.
\end{abstract}

%% file: introduction.tex
\section{Introduction}

The discovery of pulsars \citep{1968Natur.217..709H} is considered to be one of the significant findings in astronomy in the last century. Pulsars are rapidly rotating (spin periods from $\sim$ ms to s), highly magnetized ($\sim 10^8 - 10^{12}$ G) neutron stars \citep{1968Natur.217..709H, 1969Natur.221...25G} that have proved to be enormously insightful to study physics in extreme environments. Even after more than half a century since the discovery of pulsars, their emission mechanism and even the distribution of the emission regions in the magnetosphere are poorly understood. Several observed pulsar phenomena such as nulling, mode changing, subpulse drifting, etc., directly probe the emission from these compact objects. Studying these phenomena can potentially provide vital clues for uncovering the physical processes that govern how pulsars work.  

At the level of individual pulses, pulsar emission typically consists of one or more distinct components known as subpulses. \citet{1968Natur.220..231D} were the first to observe the phenomenon of drifting subpulses, whereby individual subpulses appear to drift in pulse longitude within the on-pulse region with time. As a result, the subpulses are observed to shift in phase systematically within the average pulse window between successive pulses. Several early studies delineated these pulse-to-pulse variations to be a characteristic property of pulsar emission \citep{1971ApJ...167..273T, 1973ApJ...182..245B}. When these individual pulses are stacked vertically to form a two-dimensional (2D) pulsestack, the subpulses tend to visually follow a systematic shift within the on-pulse region. This systematic shift resembles a set of discrete diagonally oriented bright regions called \textit{drift bands}. The drift bands may undergo amplitude and phase modulation with time, thereby changing the subpulse intensity or the phase at which it occurs.   

The observations of a periodic subpulse modulation suggested the presence of regularly spaced sub-beams that rotate progressively around the magnetic axis of the star \citep{doi:10.1146/annurev.aa.10.090172.002235}. Subsequently, a detailed explanation was presented in terms of the polar gap theory of \citet{1975ApJ...196...51R}. This theory suggested that the pattern of radio waves that ultimately escape the magnetosphere is the emission signature of a set of discrete, localised pockets of quasi-stable electrical activity called \textit{sparks} that exist very close to the pulsar surface. The sparks are the sites where the vacuum at the polar cap discharges into electron-positron pairs, which leads to an avalanche of particles streaming through the magnetic field lines and producing curvature radiation. These sparks move about the magnetic axis due to an \textbf{E} $\times$ \textbf{B} drift, in an arrangement resembling a fairground "carousel". As a result, the geometry of the emission beam directly reflects a configuration of sparks, with the discrete beam associated with an individual spark event called a "beamlet". The location of these sparks on the polar cap determines the geometrical pattern of the instantaneous subpulses within a pulsar's integrated pulse profile. The carousel rotation rate, $P_4$, is usually different from that of the pulsar period. Such a model result in observing a different intensity pattern with each rotation as the observer's line of sight cuts through a slightly rotated carousel, producing the observed drifting behaviour. The vertical separation between driftbands at a given longitude is $P_3$, which is measured in units of $P_1$, the pulsar rotation period. Therefore, $P_3$ is a measure of time after which a spark will return at a particular longitude. The horizontal distance between any two subpulses within a given pulse is measured as $P_2$, in units of longitude. This quantity directly translates to the separation between sparks that cross the observer's line of sight. 

While theorists have investigated the phenomenon of subpulse drifting for years, there are still multiple questions that are yet to be answered. Even though the carousel model has been scrutinised thoroughly over the years, there are some outstanding issues with the quantitative predictions of this theory.  For instance, the carousel model relies on the curvature radiation mechanism for explaining the coherent radiation, which is difficult to justify from our current knowledge of plasma physics \citep{2017RvMPP...1....5M}. It is also not clear to what extent the rotation frequency of the star aliases the observed drift rates. The model may require modifications or extensions to explain the observed phenomena like drift mode switching \citep{1970ApJ...162..727H}, nulling \citep{1970Natur.228...42B}, bi-drifting \citep{2004ApJ...616L.127Q}, etc. In essence, subpulse drifting is a vital phenomenon in understanding the physics behind the radio emission mechanism in pulsars \citep{1986ApJ...301..901R} and is known to be prevalent among a substantial fraction ($\sim$ 50 \%) of known pulsars \citep{2007A&A...469..607W}.  

Though there has been significant progress in the last 50 years towards understanding the pulsar emission process, the complex nature of pulsar emission, the exact location of emission, distribution of pulse emitting region, etc., remains poorly understood. The advent of wideband instruments and faster processing techniques brings new opportunities for further detailed investigation of the subpulse drifting phenomenon. The increased sensitivity of radio telescopes provides a unique opportunity for in-depth observations of the single pulse behaviour of pulsars with much finer time and frequency resolution. 

Even though the observational investigation of pulsar emission phenomenology has been an active area of research for the past several decades, many questions remain open. In this study, we used the upgraded Giant Meterwave Radio Telescope (uGMRT) in a dual-band phased array mode to simultaneously cover the frequency range of 300-750 MHz to study the single pulse behaviour of PSR J1822$-$2256 (B1819-22). The focus of this paper is a single pulses study of J1822$-$2256, which is a relatively less studied, bright, long period ($P_1$ = 1.874 s) pulsar with a dispersion measure (DM) of 121.2 pc cm$^{-3}$. As with several other pulsars, J1822$-$2256, is known to switch between a few different emission modes over the timescale of a single pulsar rotation (e.g., \citealp{2005MNRAS.357..859R}, \citealp{2017ApJ...836..224M}), changing the longitude positions of subpulses from one pulse to another. PSR J1822$-$2256 shows distinct drift modes, often characterised by an abrupt change in the drift rate along with amplitude modulation. The subpulse drifting properties of this pulsar have been previously investigated by \citet{2006A&A...445..243W, 2007A&A...469..607W, 2009A&A...506..865S, 2017A&A...604A..45N, 2018IAUS..337..348J, 2018MNRAS.476.1345B}. The mode changing varies the value of $P_3$ over different modes of emission, implying a change in either the rotation speed of the carousel or in the distribution or number of sparks. Previous studies \citep{2006A&A...445..243W, 2009A&A...506..865S, 2018MNRAS.476.1345B} have reported the existence of multiple drifting modes, including two modes, corresponding to $P_3$ values of $16\,P_1$ and $6\,P_1$ respectively, and a non-drifting mode. In addition, \citet{2018MNRAS.476.1345B} also found a transitional mode at $14\,P_1$, that the pulsar sometimes exhibits before entering in mode A. \citet{2018IAUS..337..348J} have found three modes of drifting corresponding to $P_3$ values of $17\,P_1$, $7.5\,P_1$, and $5\,P_1$. This pulsar also has a wide profile with about an 8\% duty cycle, and thus the line-of-sight of earth-based observer samples a large section of the polar cap.

Another intriguing phenomenon is pulse nulling, whereby no detectable emission is seen above the telescope's sensitivity threshold for some time, after which the pulsar emission reappears suddenly to its normal state. The definition of nulling clearly depends upon the sensitivity of the telescope. With the wider bandwidth and higher sensitivity of our data, it is possible to search for a low-level (and persistent) emission during the nulls. To date, nulling \citep{10.1111/j.1365-2966.2007.11703.x, 10.1111/j.1365-2966.2012.21296.x} has been observed in a large number of pulsars. The pulsar J1822$-$2256 is also studied to null with a nulling fraction of $\sim$ 10$\pm$2\% \citep{2017A&A...604A..45N, 2018IAUS..337..348J, 2018MNRAS.476.1345B}.

In this paper, we present a detailed investigation of single-pulse properties of  PSR J1822$-$2256 at 300-750 MHz using the uGMRT. The remainder of the paper is organised as follows. The observations and data-processing are summarised in section \ref{sec:obs} and \ref{sec:dataprocess}. In section \ref{sec:modechange}, we re-visited the subpulse drifting modes for this pulsar in light of the high sensitivity multi-frequency data. Other characteristics, including brief analysis of $P_2$, $P_3$, and nulling are described in section \ref{sec:p3dist} to \ref{sec:p2dist}. Section \ref{sec:freqdep} describes to the frequency evolution of the subpulse behaviour. A discussion based on the results of our study is presented in Section \ref{sec:discussion}, including the relationship between different modes and the carousel geometry. Finally, our conclusions are presented in Section \ref{sec:conclusion}.

%% file: observations.tex
\section{Observations} \label{sec:obs}

The  Giant Metrewave Radio Telescope (GMRT) is a radio interferometric array with 30 antennas, each of 45-meter diameter, configured in a Y shape over an area of 28 km$^2$ \citep{1991CuSc...60...95S}. In a recent upgrade, the system was equipped with wide-band receivers and digital instrumentation to provide a near-seamless coverage in frequency from 120 MHz to 1600 MHz \citep{2017CSci..113..707G, 2017JAI.....641011R}, and is now called the upgraded GMRT or uGMRT. We used the phased array mode of the uGMRT, where the incoming signals from all antennas are coherently combined for maximum sensitivity for pulsar observations. For our observations, we used 13 antennas for Band 3 (300-500 MHz) and the same number for Band 4 (550-750 MHz), where we recorded the total intensity data. Observations were made at two epochs, separated by $\sim$ 10 days, simultaneously at both frequency bands. At Epoch 1, we recorded 1.5 hours of data, containing $\sim$ 2900 pulses. Epoch 2 was $\sim$ 53 minutes long and recorded $\sim$ 1700 pulses. The observation details are summarised in Table \ref{tab:obs}.  

\input{table1.tex}

The data were recorded with a time resolution of 327.68 $\mu$s, and a frequency resolution of 48 kHz, i.e., 4096 frequency channels over the 200 MHz observing band. Given the pulsar DM of 121.2 pc cm$^{-3}$ for this pulsar, the dispersive smearing is 116 $\mu$s across each channel at the highest observation frequency (750 MHz) and 1.82 ms at the lowest (300 MHz) observation frequency. These numbers imply an effective resolution of 116 $\mu$s and 1.82 ms at the highest and lowest frequency channel, respectively.

The data were integrated to 1.826 ms, with 1024 bins across the pulsar period. The resulting data were written in the filterbank format and subsequently converted to single pulse archives by processing using the {\tt DSPSR} package \citep{2011PASA...28....1V}. The single pulse files were then frequency scrunched and combined using the routines from {\tt PSRCHIVE} \citep{2004PASA...21..302H}. Finally, the frequency scrunched single pulse sequence file was manually searched for radio frequency interference (RFI) using the interactive RFI zapping subroutine {\tt pazi} of {\tt PSRCHIVE}. The RFI-excised file was then converted into an ASCII format that contained the pulse time series and was used for all subsequent analyses.

%% file: table1.tex
\begin{table}
\caption{Details of observation of J1822-225 using uGMRT}
\label{tab:obs}
\begin{tabular}{l|c|c|c|c|}
\hline
\multirow{2}{*}{\textbf{}}                     & \multicolumn{2}{c|}{Band 3}    & \multicolumn{2}{c|}{Band 4}    \\
                                               & Epoch - 1 & Epoch - 2 & Epoch - 1 & Epoch - 2 \\ \hline
\multicolumn{1}{|l|}{Obs. Length (s)} & 5400               & 2880               & 5400               & 2880               \\ \hline
\multicolumn{1}{|l|}{MJD}             & 59010                  & 59020                 & 59010                  & 59020                 \\ \hline
\multicolumn{1}{|l|}{Singlepulse S/N} & 33.29                  & 8.62                  & 2.99                  & 2.82                  \\ \hline
\multicolumn{1}{|l|}{Freq. Res. (kHz)}       & 48.828               & 48.828               & 48.828               & 48.828               \\ \hline
\multicolumn{1}{|l|}{Time Res. ($\mu$s)}       & 836.82               & 836.82               & 373.29               & 373.29               \\ \hline
\multicolumn{1}{|l|}{No. of Antennas}     & 13                 & 13                 & 13                 & 13                 \\ \hline
\multicolumn{1}{|l|}{No. of Pulses}       & 2879               & 1704               & 2879               & 1704               \\ \hline
\end{tabular}
\end{table}

%% file: analysis.tex
\section{Analysis}

The wide bandwidth capability of the uGMRT provides greater sensitivity for single pulse studies. The improvement enables approximately 6$\times$ increase in the usable bandwidth, and thence a sensitivity boost by over a factor of two for any single pulse work. The simultaneous multi-frequency observations allowed us to study the subpulse drifting, mode changing, and the nulling phenomenon over an extended (and nearly continuous) frequency range from 300 to 750 MHz.

\begin{figure*}
    \includegraphics[width=\columnwidth,trim={1.5cm 0cm 1.5cm 1cm},clip]{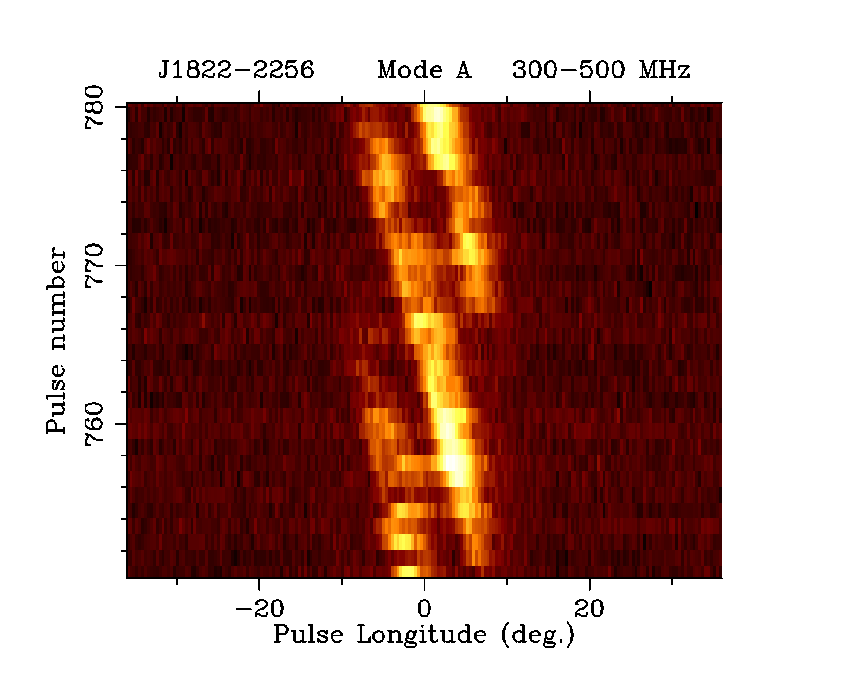}
    \includegraphics[width=\columnwidth,trim={1.5cm 0cm 1.5cm 1cm},clip]{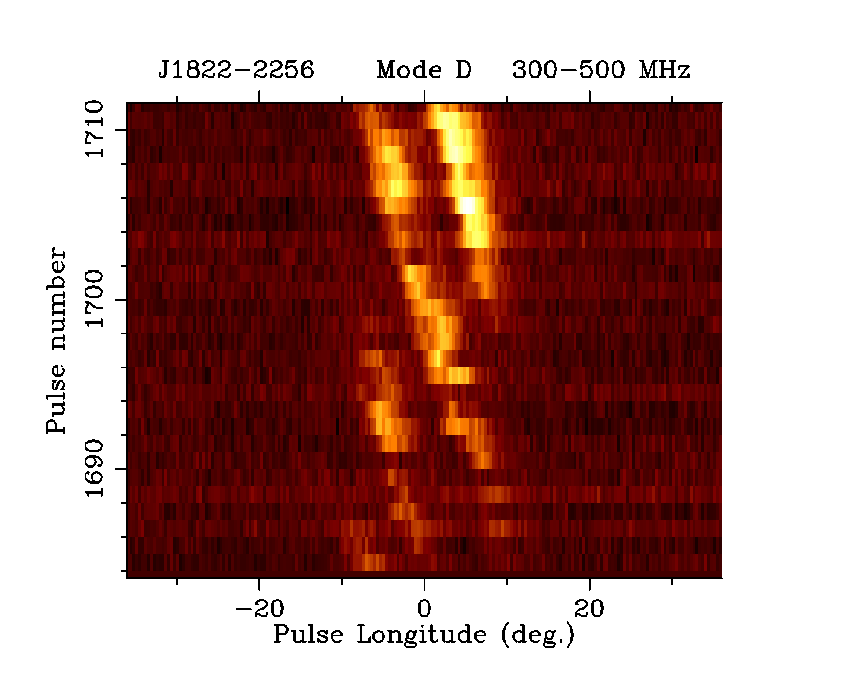}
	\includegraphics[width=\columnwidth,trim={1.5cm 0cm 1.5cm 1cm},clip]{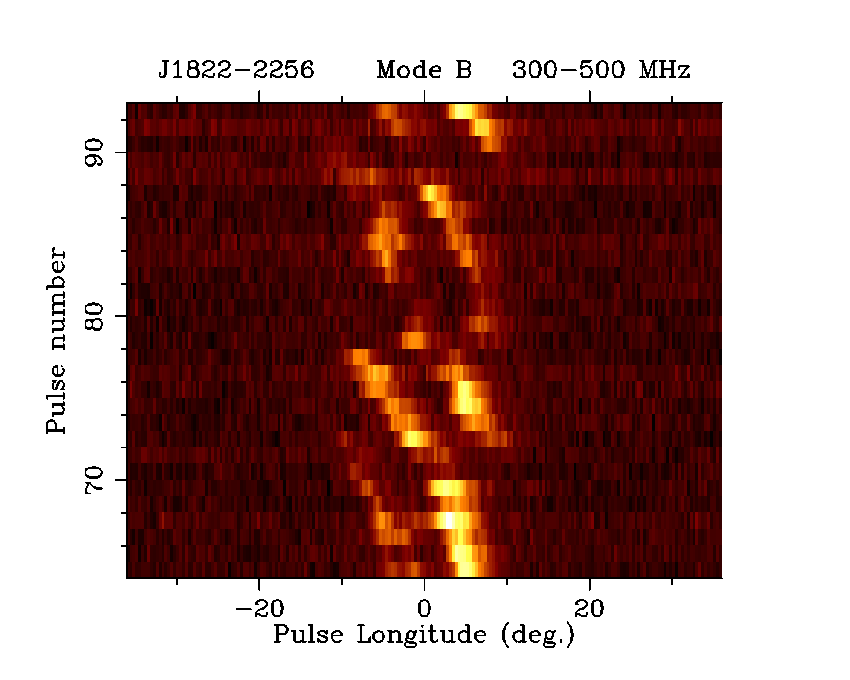}
	\includegraphics[width=\columnwidth,trim={1.5cm 0cm 1.5cm 1cm},clip]{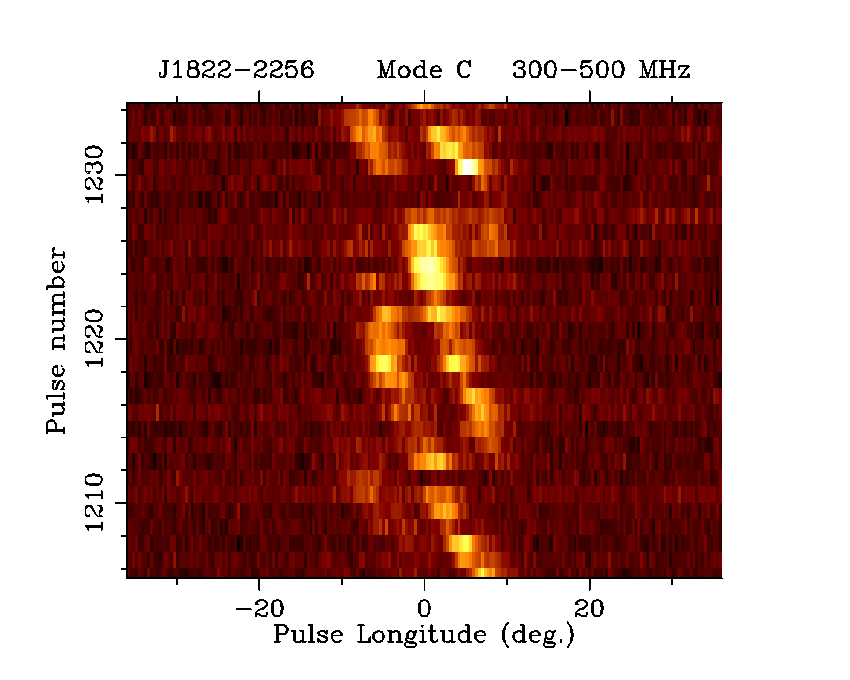}

    \caption{2D-pulsestacks displaying the subpulse drifting modes of emission observed for J1822-2256. The modes A, B, C, and D are arranged anti-clockwise. Here the $x-$axis shows the pulse phase, whereas the $y-$axis is the pulse number. The figure clearly shows drift patterns and their distinct behaviour representing the four different drift modes.}
    \label{fig:modeps}
\end{figure*}

The single pulse sequence reveals the presence of multiple emission modes with distinct characteristics. Examples of two-dimensional (2D) pulse stacks are shown in Fig. \ref{fig:modeps}. Here the $x$-axis represents the pulse phase, where the pulse longitude range $0$-$360^\circ$ corresponds to the phase range $0$-$1$. The four panels represent the four distinct drift modes (A, B, C, and D) that we observed for this pulsar. Each of the sub-pulse drifting modes is further described in Section \ref{sec:modechange}. The high sensitivity single pulse observations allowed us to identify the emission modes visually. Furthermore, the pulsar also exhibits significant variation in its drifting behaviour, modulating the intensity and the apparent position of the subpulses. The following sections describe the analysis performed to study the sub-pulse drifting behaviour of the pulsar.

\begin{table}
	\centering
	\caption{Drift mode measurements of J1822$-$2256}
	\label{tab:modes}
	\begin{tabular}{lcccr} % four columns, alignment for each
		\hline
		\hline
		Mode & \begin{tabular}[c]{@{}c@{}}Average Burst \\ Length\end{tabular} & \begin{tabular}[c]{@{}c@{}}Mean P3 \\ (P1)\end{tabular} & \textbf{\begin{tabular}[c]{@{}c@{}} \textbf{$\sigma_{P_3}^{a}$} \end{tabular}} & \begin{tabular}[c]{@{}r@{}} $O^b$ \\ (\%)\end{tabular} \\ \hline
		\hline
		A & 83 & 17.92 $\pm$ 0.02 & 2.0 & 46\\
		B & 10 & 5.84 $\pm$ 0.16 & 3.6 & 19\\
		C & 15 & 7.97 $\pm$ 0.15 & 2.5 & 17\\
		D & 22 & 14.09 $\pm$ 0.02 & 0.9 & 6\\
		Pseudo Null & 2 & - & - & 5\\
		Real Null & 8 & - & - & 5\\
		Unknown$^c$ & 10 & - & - & 2 \\
		\hline
	\end{tabular}
	\begin{tablenotes}
        \small
        \item $^a$ Standard deviation of $P_3$ in each mode        
        \item $^b$ \% of occurrence of the particular drift mode
        \item $^c$ Unknown: pulses that could not be classified in one of the four modes of emission.
    \end{tablenotes}
\end{table}

%%%%%%%%%%%%%%%%%%%%%%%%%%%%%%%%%%%%%%%%%%%%%%%%%%%%%%%%%%%%%%%%%%%%%

\subsection{Mode change boundaries and characterising $P_3$ values} \label{sec:dataprocess} 

To determine the mode change boundaries, we employed the Phase Averaged Power Spectrum (PAPS) method described in \citet{2005A&A...440..683S}. We first visually selected a trial sequence of pulses that roughly belonged to one kind of specific drifting sequence. For each of these pulse sequences, we obtained the distribution of intensity values across the pulse number at a fixed phase. We then performed a Fourier transform of the intensity distribution at each phase to obtain the power spectrum for each phase bin within the on-pulse region. Finally, all the power spectra were averaged over phase, giving a PAPS from 0 to 0.5 cycles per period. 

\begin{figure}
    \centering
    % trim={<left> <lower> <right> <upper>}
    \includegraphics[width=\columnwidth]{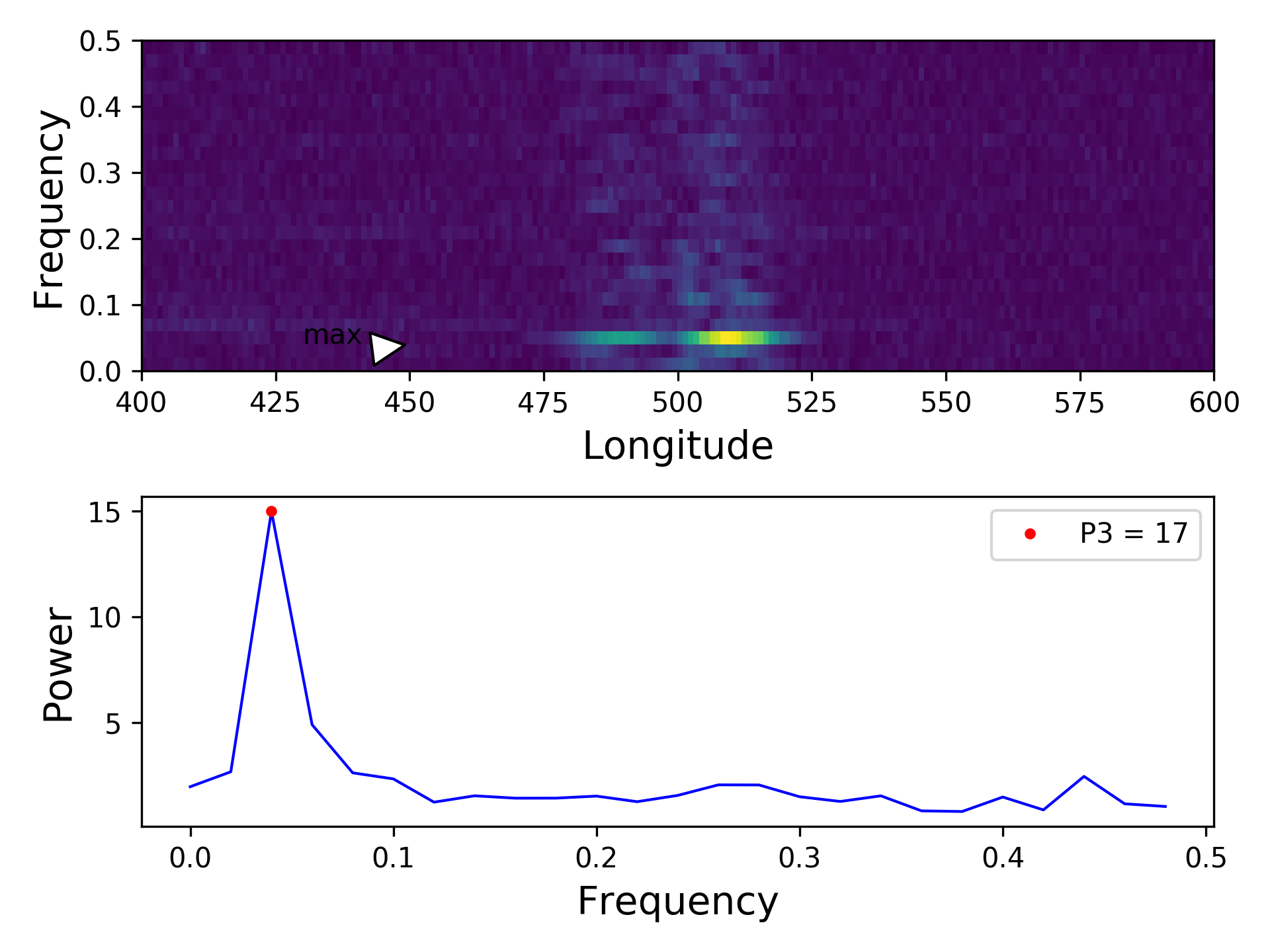}
    \caption{LRFS and PAPS for the mode A pulse sequence shown in Fig. \ref{fig:modeps}. The top panel shows the Longitude Resolved Fluctuation Spectra (LRFS) of the sequence. The bottom panel shows the phase/longitude averaged value of the LRFS, i.e. PAPS. The maximum intensity mark on the top panel corresponds to the peak in the bottom panel. The inverse of this peak frequency position is $P_3$.}
    \label{fig:papsA}
\end{figure}

\begin{figure*}
	\includegraphics[width=\textwidth]{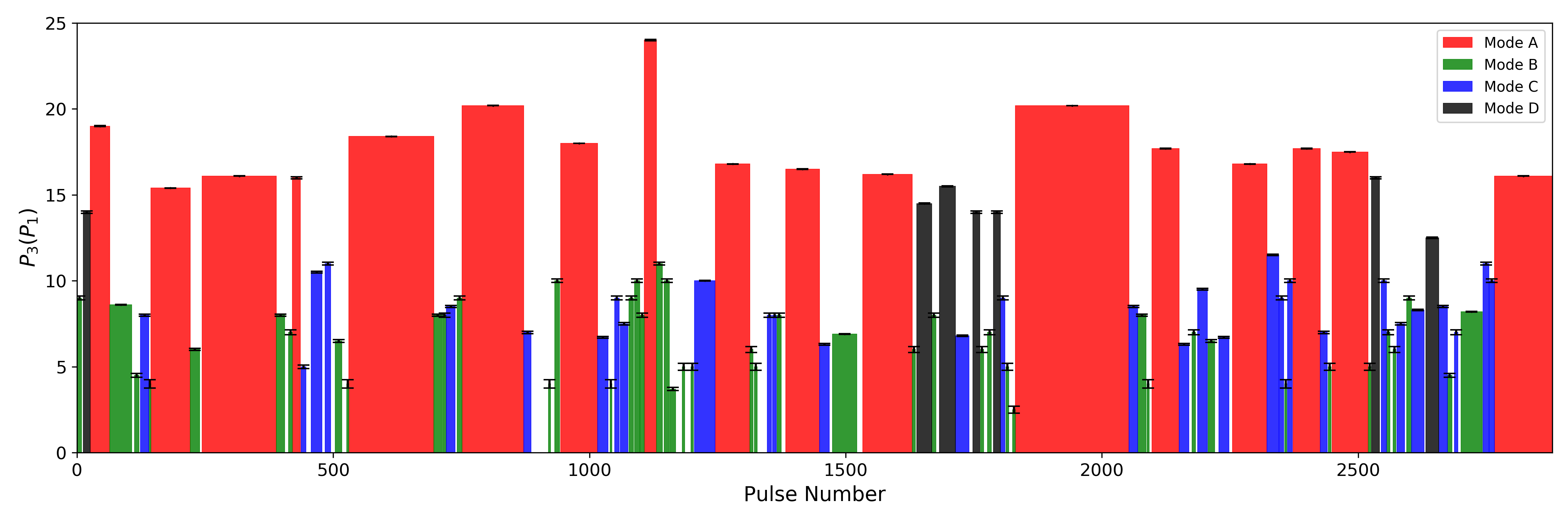}\\
	\includegraphics[width=\textwidth]{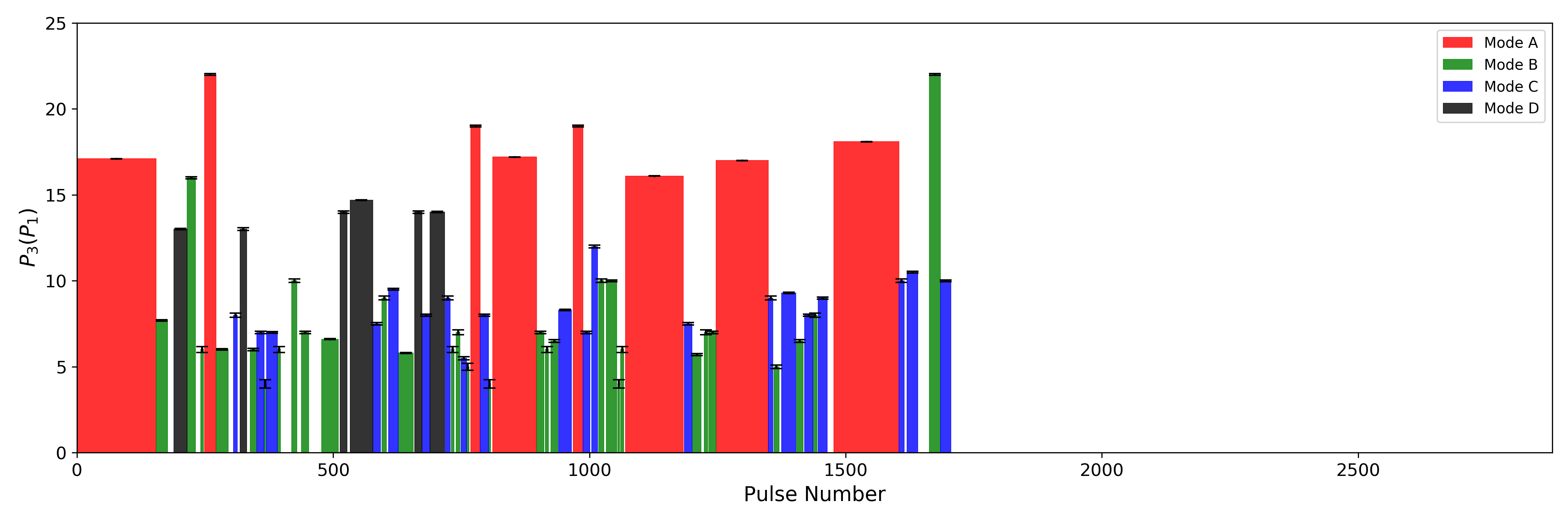}
    \caption{A plot of $P_3$ values for Epoch 1 (top) and Epoch 2 (bottom) of J1822$-$2256 as calculated using the Phase Averaged Power Spectrum (PAPS). The x-axis shows pulse number and the y-axis shows $P_3$ in units of $P_1$. A single block represents a continuous stretch of pulses showing one kind of subpulse drifting mode. The error bars are defined as the inverse of the number of pulses in the respective sequence.}
    \label{fig:PAPS}
\end{figure*}

An example is given in Fig. \ref{fig:papsA}, showing the PAPS for a pulse sequence corresponding to mode A. Whenever a peak was found in the PAPS, the beginning and end of the pulse sequence were adjusted to get the highest signal-to-noise ratio (S/N) for the peak. The S/N of the peak was calculated by dividing the peak value of the PAPS with the root mean square value of the rest of the PAPS. The start and end pulse numbers of the sequence were then visually inspected to check their coherence with the values calculated from PAPS. We used the reciprocal of the frequency of the centre of the peak in the PAPS to calculate the $P_3$ value for the respective drift sequence. The resulting $P_3$ values of each sequence for Epoch 1 and Epoch 2 data are shown in Fig.~\ref{fig:PAPS}. The frequency resolution, given by the inverse of the number of pulses in the sequence, was taken as the error on the peak position.

%%%%%%%%%%%%%%%%%%%%%%%%%%%%%%%%%%%%%%%%%%%%%%%%%%%%%%%%%%%%%%%%%%%%%

\subsection{Modes of Emission} \label{sec:modechange}

\begin{figure*}
	\includegraphics[width=0.9\columnwidth]{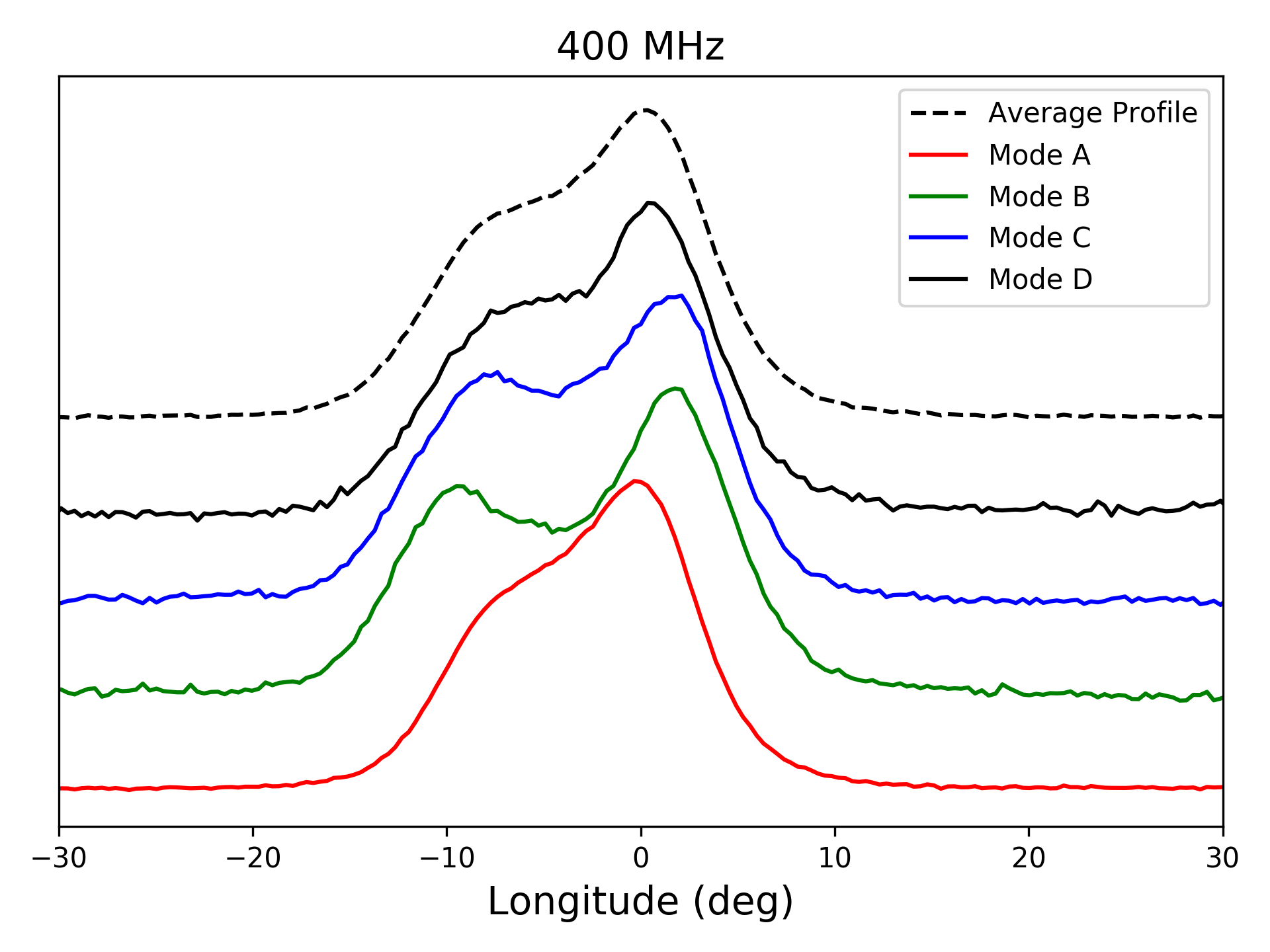}
	\includegraphics[width=0.9\columnwidth]{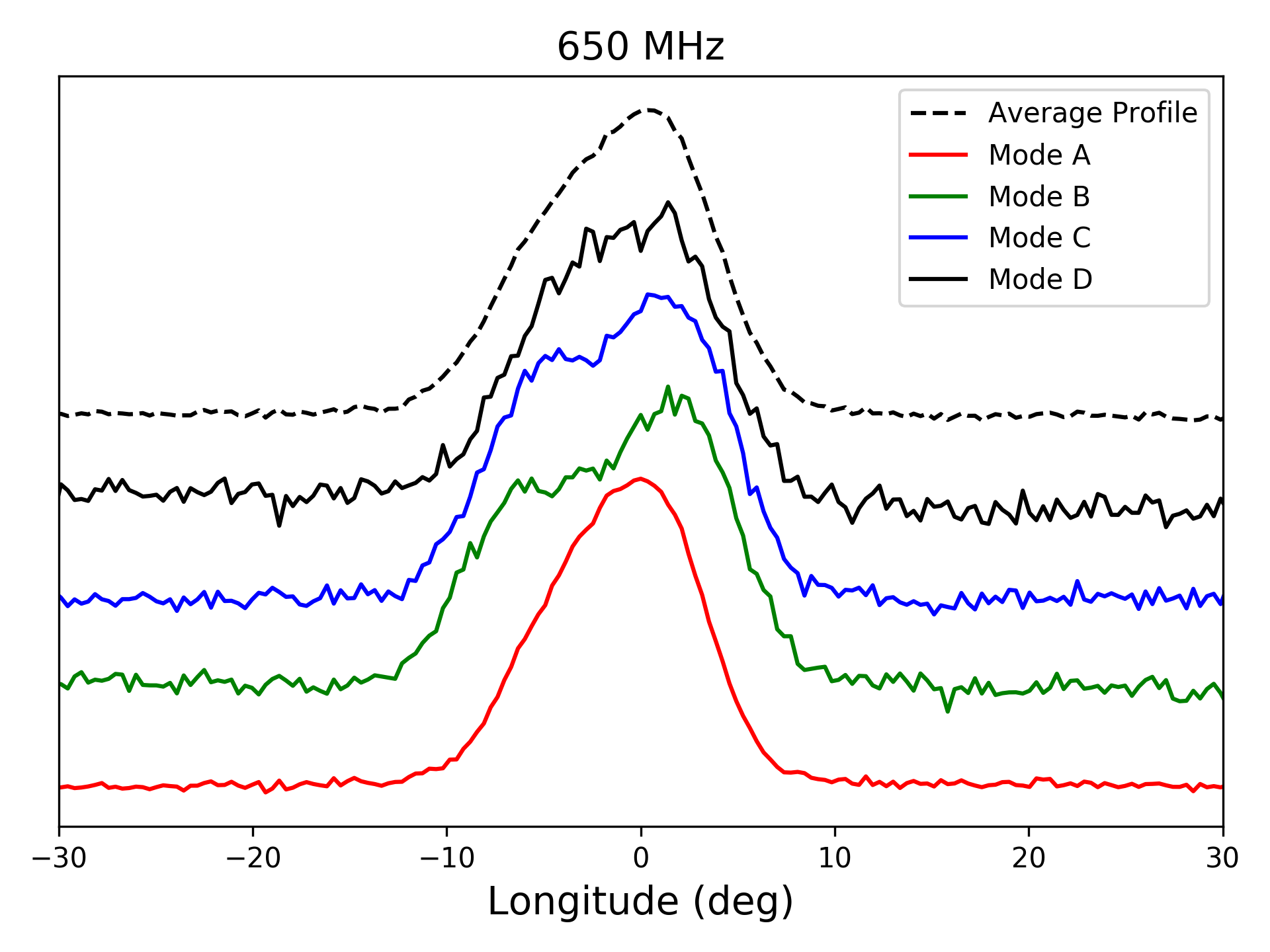}
    \caption{Average profiles of different modes of emission of J1822$-$2256 at Band 3 (300-500 MHz) and Band 4 (550-750 MHz). The black dashed lines show the average profile, and the rest show colour-coded average profile for each mode. The profiles are given arbitrary constant shifts along the y-axis for display purposes. Apart from the variation of the profiles across the different modes, the frequency dependent behaviour of the profiles in each mode can also be observed here.}
    \label{fig:modeprof}
\end{figure*}

Previous studies of PSR J1822$-$2256 have revealed multiple distinct modes of emission. Our analysis suggests the presence of four subpulse drifting modes, including some new features that were not reported earlier. We used the $P_3$ value as well as the average drift profile as the primary criteria for identifying these modes. A summary of the drift mode statistics is given in Table \ref{tab:modes}. Fig. \ref{fig:modeprof} shows the average profile of all the categorised modes at both the observed frequency bands along with the full pulsar profile in each band. The centred on-pulse window lies roughly in the pulse longitude range -20$^{\circ}$ $\lesssim$ $\phi$ $\lesssim$ +12$^{\circ}$ for Band 3 (300-500 MHz) and -15$^{\circ}$ $\lesssim$ $\phi$ $\lesssim$ +10$^{\circ}$ for Band 4 (550-750 MHz), where the longitude $\phi = 0 ^{\circ}$ is the centre of the profile where the maximum average flux density is maximum. 

\subsubsection{Mode A}
 Mode A shows prominent drift bands with a characteristic drifting behaviour (Fig. \ref{fig:modeps}) with a mean $P_3$ value of 17.9 $P_1$. The $P_3$ values of different occurrences show a spread of $\sim 2\;P_1$ from the mean $P_3$. This mode was previously reported in the literature \citep{2007A&A...469..607W, 2018IAUS..337..348J, 2018MNRAS.476.1345B}. Our observations suggest an occurrence fraction of 46\%. The average profiles for mode A at Band  3 and Band  4 are shown in Fig. \ref{fig:modeprof}, and can be compared to the integrated profile of other modes.

\subsubsection{Mode B}
The pulsar frequently transitions to mode B, which is present for approximately 19\% of the time during our observations. This is likely to be similar to mode B as mentioned in \citet{2018MNRAS.476.1345B}; however, their study showed this mode to be a non-drifting mode. Our analysis shows mode B drifting with a mean $P_3$ value of $5.8\;P_1$. \citet{2018IAUS..337..348J} find a similar mode with $P_3$ of $5\;P_1$ which they refer to as mode C. The $P_3$ values of different occurrences of this mode show a wide spread of $\sim 3.6\;P_1$ from the mean $P_3$. In our case, mode B occurs in short bursts, with frequent nulls (Fig. \ref{fig:modeps}). We categorised all pulse sequences as mode B, where the drift modes had discontinuities in the drift bands or where the mode length was less than ten pulses. In these cases, a full drift mode is not observed; however, partial drift bands are visible with the characteristic drift rate of mode B. The average profile for mode B (Fig. \ref{fig:modeprof}) shows a double-peaked structure with widely separated components due to the broken features.

In some cases in the pulse train, the driftbands had a few instances that are marked by an apparent lack of emission in either the middle or the latter half of the drift sequence, leaving the rest of the sequence relatively intact, whereas, in other cases, the drift sequences exhibited a greater suppression of the intensity, such that it becomes relatively difficult to discern a drift band, even though the drifting feature itself is present (Fig. \ref{fig:modeps}). The modes of the latter type were classified as mode B. 

\subsubsection{Mode C}
We measure a slightly different mean $P_3$ for this mode; however, it could possibly be same as the mode C mentioned in \citet{2018MNRAS.476.1345B}, as the values are consistent within the error bars. \citet{2018IAUS..337..348J} also mention a drifting mode at $7.5\;P_1$ which they call mode B. In our case, mode C shows a mean $P_3$ of $\sim 8\;P_1$ with different occurrences showing a spread of $\sim 2.5\;P_1$ from the mean $P_3$. In mode C, the drift sequences showed occasional lack of detectable emission in the drift band (Fig. \ref{fig:modeps}). In all such cases, the lack of emission is typically seen in the centre or the top part of the drift sequence. Unlike mode B, it is easier to discern and construct a well-connected drift band for mode C, even if a detectable emission was absent for some pulses. It is also possible that mode B and C are related, and the drift bands become irregular as they go to lower $P_3$ values. The corresponding average profile for mode C (Fig. \ref{fig:modeprof}) shows two components, similar to that of mode B, but with much closely separated profile components and a smaller profile width.

\subsubsection{Mode D}
We observed another mode -- mode D -- which is present only for about 6\% of the observing duration. This mode has no visible breaks in the drift sequence. However, it shows occasional intensity variations over the drift band, usually brighter in the leading part of the profile (Fig. \ref{fig:modeps}). It displays a characteristic drifting nature similar to mode A, but a mean $P_3$ value of $\sim 14\;P_1$ and a different average profile (Fig. \ref{fig:modeprof}). The $P_3$ values of different occurrences of this mode show a spread of $\sim 1\;P_1$ from the mean $P_3$. \citet{2018MNRAS.476.1345B} found a transitional mode at a similar $P_3$ value, which occurred mostly when the pulsar was transitioning from mode A to any other mode. However, we choose to call this type of pulse sequence as mode D because of the characteristic difference between the two modes, i.e., most of the occurrences of mode D for our case were preceded by a null. Out of the 13 occurrences of mode D in our observations, only two were found to follow immediately after another mode, namely mode B or C. \citet{2018IAUS..337..348J} do not report any mode occurring at this $P_3$.

\subsection{$P_3$ distribution} \label{sec:p3dist}
\begin{figure}
	\includegraphics[width=\columnwidth]{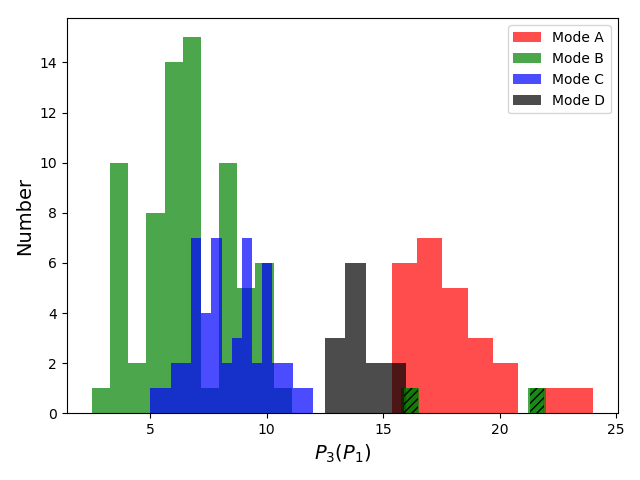}
    \caption{Plot of the distribution of the $P_3$ values for J1822$-$2256 as calculated by PAPS. The different modes are displayed with different colours. The green-dashed data are part of mode B but are due sequences which were affected with RFI and also showed apparent lack of emission within the drift band.}
    \label{fig:P3full}
\end{figure}

Fig.~\ref{fig:P3full} shows the distribution of $P_3$ values for the full observations, where colour-coded values represent different drift modes. The $P_3$ values span a large range of values, with mode B showing the smallest value, in which fewer pulses show multiple short-duration bursts. Mode A exhibits the largest values for $P_3$ with fewer longer duration bursts and a larger occurrence fraction. The two mode B occurrences at $P_3$ $\sim$ 16 and 22 $P_1$ (Fig. \ref{fig:P3full}, shown in green-dashed) are due to sequences that were affected with RFI and also showed no detectable emission within the drift band at times. Since these show the typical characteristics of mode B, i.e., an apparent lack of emission in some parts of the drift band and a wide average profile, the two occurrences were included in mode B. We also searched for the possibility of a correlation between the mode lengths and the $P_3$ values for each mode. However, our analysis showed no such significant correlation.

%%%%%%%%%%%%%%%%%%%%%%%%%%%%%%%%%%%%%%%%%%%%%%%%%%%%%%%%%%%%%%%%%%%%%

\subsection{Nulling behaviour of J1822$-$2256}

Individual pulses tend to vary in intensity on a variety of timescales, and this is attributed to intrinsic emission processes. A notable phenomenon is pulse nulling, whereby there is no detectable emission above the sensitivity limit of a given observation. In our analysis, we found two different kinds of nulls for J1822$-$2256. Each of the single pulses were visually inspected to check for any emission feature within the on-pulse longitude ranges. Pulses that did not show a detectable emission feature in the on-pulse region were categorised as null. The pulsar shows regular broadband nulls for about 5\% of the total observing time, with an average null length of eight pulses. The averaged profile constructed from these nulls shows a noise-like behaviour at both the observing frequencies.  

\begin{figure}
	\includegraphics[width=\columnwidth]{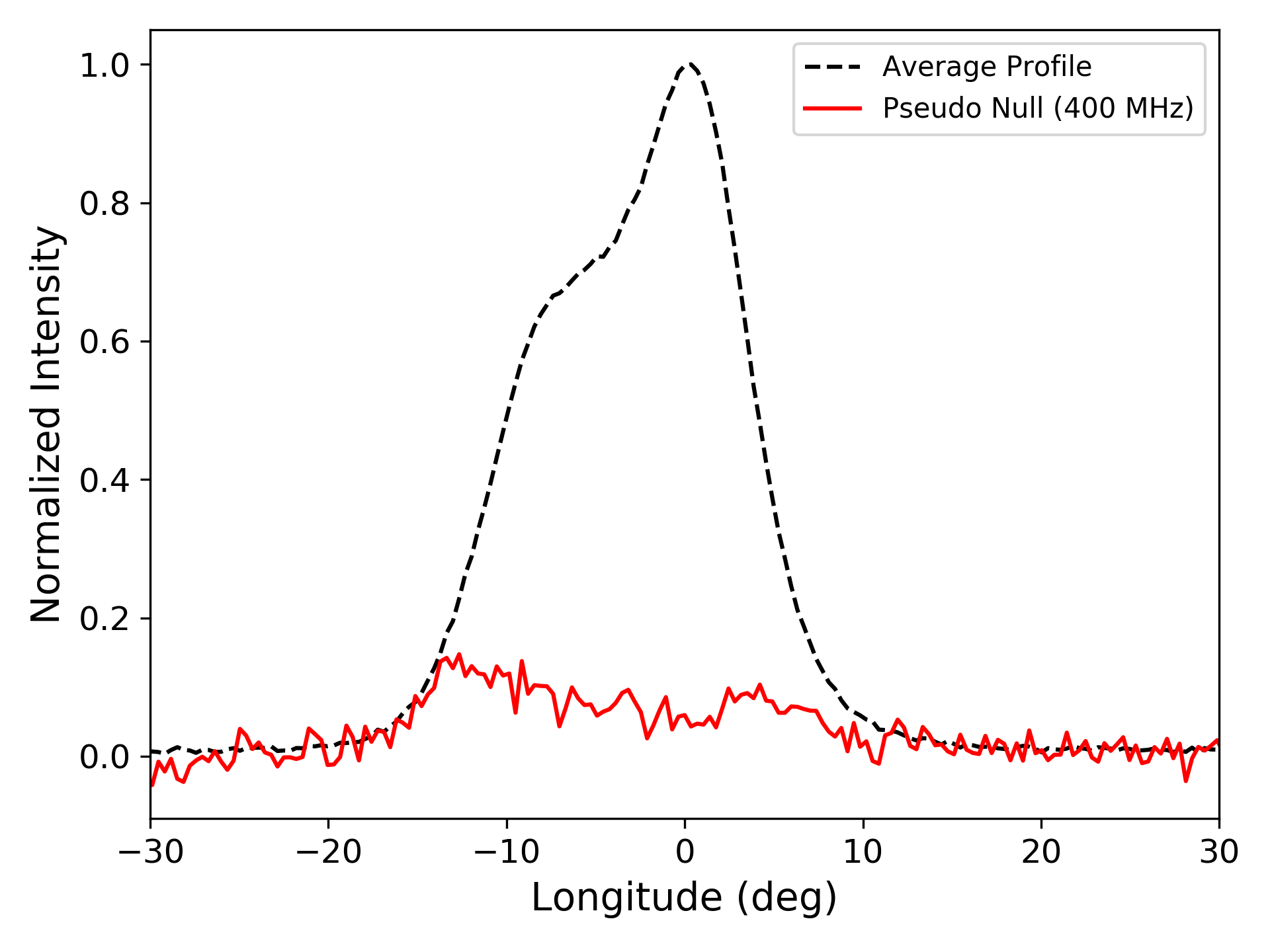}
	\includegraphics[width=\columnwidth]{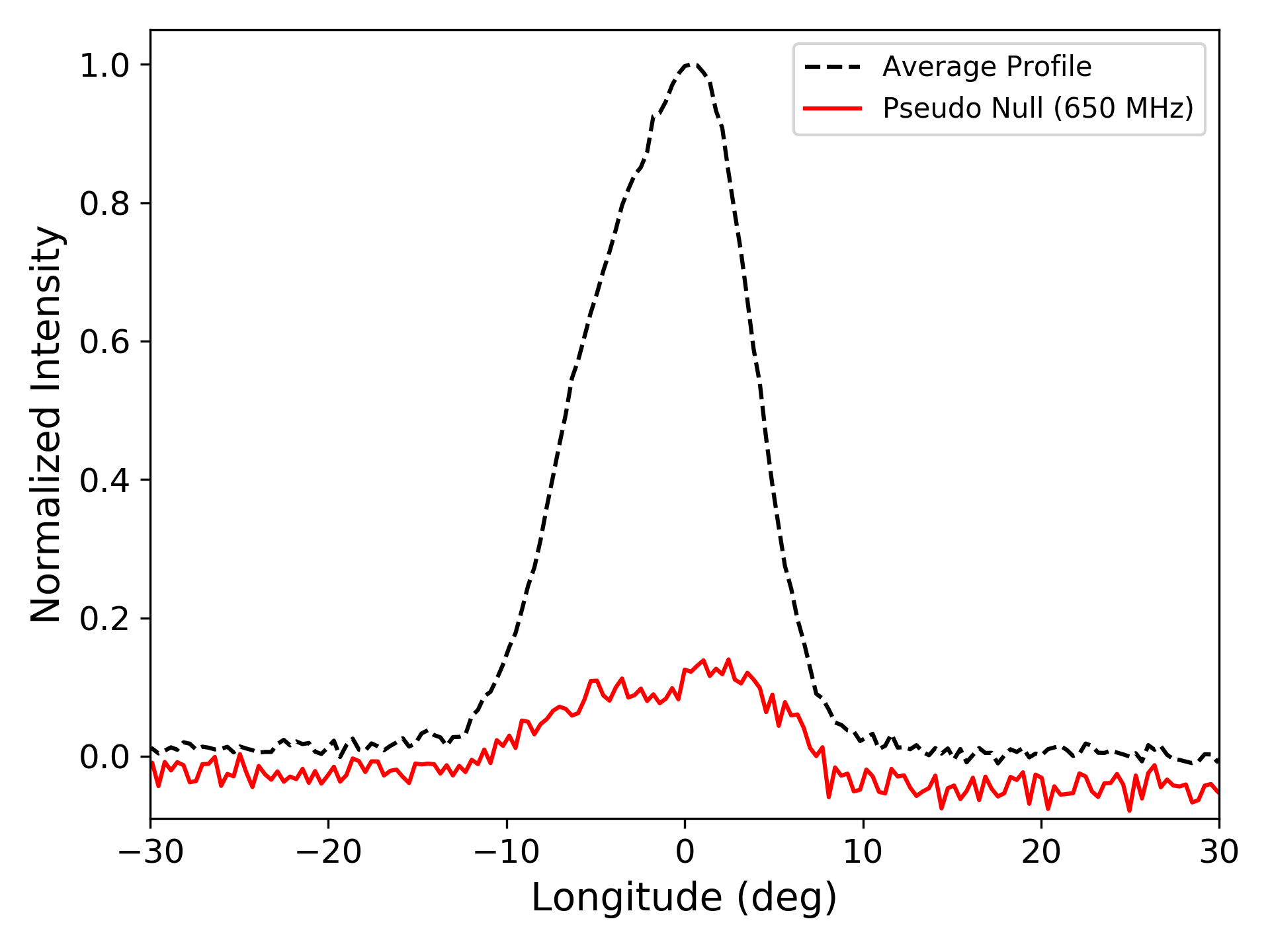}
    \caption{Profile of pseudo nulls vs. full pulse profile for Band 3 (top) and Band 4 (bottom).}
    \label{fig:pseudonull}
\end{figure}

The null occurrences which were less than or equal to two pulses in length were separately combined to form an average profile. This profile shows discernible low-level emission in both the observing bands, as shown in Fig. \ref{fig:pseudonull}. Pulses of this kind amount to 4\% of the observing duration, and we refer to them as \textit{pseudo nulls}. On a single pulse level, these \textit{pseudo nulls} show precisely the same behaviour as the real nulls. These pseudo nulls mostly occur around modes B and C and occasionally before transitioning to a different mode. 

%%%%%%%%%%%%%%%%%%%%%%%%%%%%%%%%%%%%%%%%%%%%%%%%%%%%%%%%%%%%%%%%%%%%%

\subsection{Characterising $P_2$} \label{sec:p2dist}

We also calculated the separation between two successive subpulses within a single pulse, which is the periodicity $P_2$. Abrupt changes in the number or position of the sparks may lead to a varying value of $P_2$. For calculating this parameter, we used the Two-Dimensional Fluctuation Spectra (2DFS; \citealp{2003A&A...410..961E}) method. For each set of pulses within a given mode, we first performed an FFT at each longitude/phase of the 2D-pulsestack and then performed an FFT for each pulse in the pulse train. This exercise gave two peaks in the resulting 2-dimensional transform. One corresponds to the periodicity in the vertical direction, i.e., for each phase ($P_3$), and another peak corresponds to the periodicity within each pulse ($P_2$). The average $P_2$ at Band 3 was calculated to be $12.8^{\circ} \pm 0.7^{\circ}$ and $13.4^{\circ} \pm 1.4^{\circ}$ at Band 4 over all modes. Given the small frequency gap between the two bands and the width of the individual bands themselves, the $P_2$ values are consistent within errors.

%%%%%%%%%%%%%%%%%%%%%%%%%%%%%%%%%%%%%%%%%%%%%%%%%%%%%%%%%%%%%%%%%%%%%

\subsection{Investigation of the frequency dependence of profile components}\label{sec:freqdep}

The wide-band nature of our data provides us the advantage over previous studies, as it also allows studying the frequency evolution of the subpulse drifting phenomenon. A frequency evolution of modal profiles is clearly seen in our observations (Fig. \ref{fig:modeprof}), whereby the average profile changes with an increase in frequency. The characteristic double-peaked structure becomes less separated at a higher frequency for each mode, and the profile becomes narrower.

Using the advantage of the large bandwidth provided by the uGMRT, we can also divide the observation bandwidth into smaller chunks (i.e., sub-bands). We split the observations into 100 MHz bandwidths centred at four central frequencies of 350 MHz, 450 MHz, 600 MHz, and 700 MHz. This exercise resulted in sub-banded data points to investigate the frequency evolution of the pulsar emission. We then studied the individual modal profiles at each frequency and examined their evolution. 

\begin{figure*}
	\includegraphics[width=\textwidth]{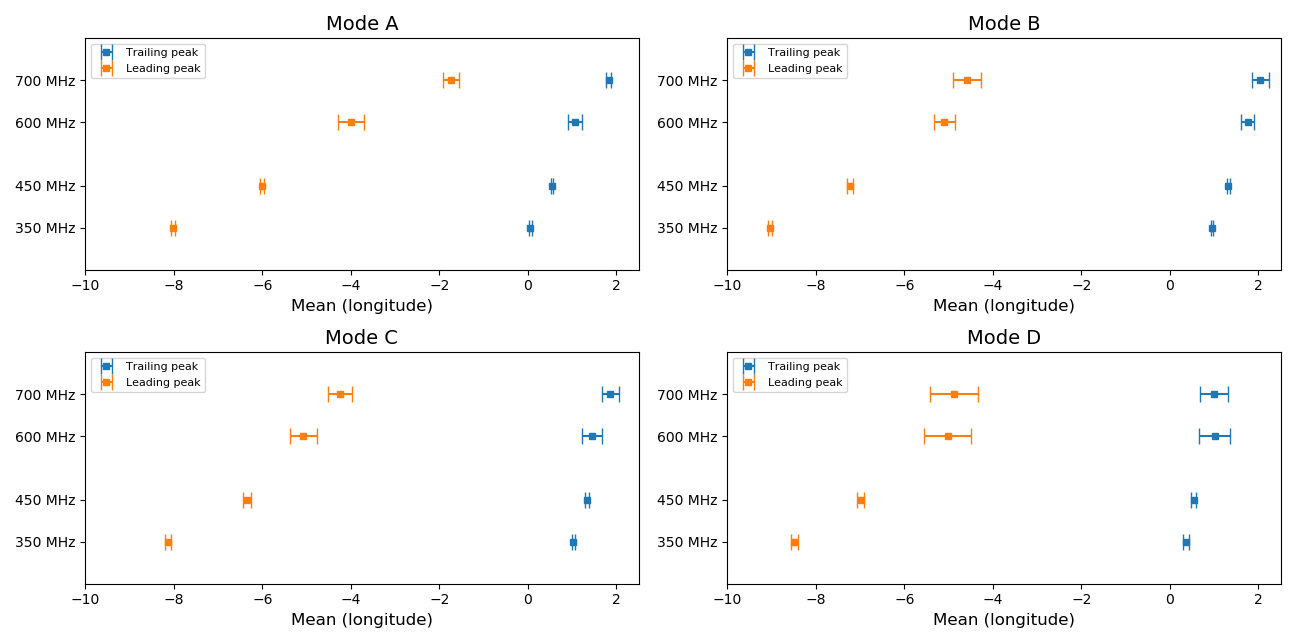}
	 \caption{Plot of mean longitude position of pulse profile components. The points on the left (orange) correspond to the leading components, whereas points on the right (blue) corresponds to the trailing component of the profile.}
    \label{fig:gfit}
\end{figure*}

We attempted to fit a combination of Gaussians to all the modal profiles, and a double Gaussian function was found to be the best model. This is a favoured model for our case because, on the single pulse level, the pulsar shows two subpulses within a given pulse, which eventually gives rise to a double-peaked profile. The frequency evolution of these two components of the double Gaussian function was examined for each mode and is plotted in Fig. \ref{fig:gfit}. The different panels correspond to the different drift modes, where the mean position of the individual Gaussian peaks (of the double Gaussian function) is plotted against the frequency. The mean positions on the left (in orange) correspond to the leading peak in the Gaussian fitted profile, whereas the ones in blue correspond to the trailing peak. A clear trend noticeable in the mean positions can be seen with increasing frequency for both components. 

\subsection{Probing the emission geometry with polarisation observations} \label{sec:polobs}

Polarimetric observations can be used to constrain the pulsar geometry and the location of radio emission. The pulsar radiation is thought to be strongly polarised either along or orthogonal to the open magnetic field lines. When the pulsar beam sweeps past the observer's line of sight, the position angle (PA) of the linear polarisation changes, resulting in an S-shaped curve. The inclination angle, $\alpha$, between the magnetic and rotational axes and the impact parameter, $\beta$, which is the smallest angle between the locus of the line of sight of the pulsar beam and the magnetic axis, determine the shape of the S-shaped curve. In the rotating vector model (RVM, \citealp{1969ApL.....3..225R}), the characteristic S-shape curve of PA vs. pulse phase is used to determine the geometrical angles for a pulsar. We use this for our analysis to infer the pulsar beam geometry. 

\begin{figure}
	\includegraphics[width=\columnwidth, trim={1.5cm 0cm 4cm 2.5cm},clip]{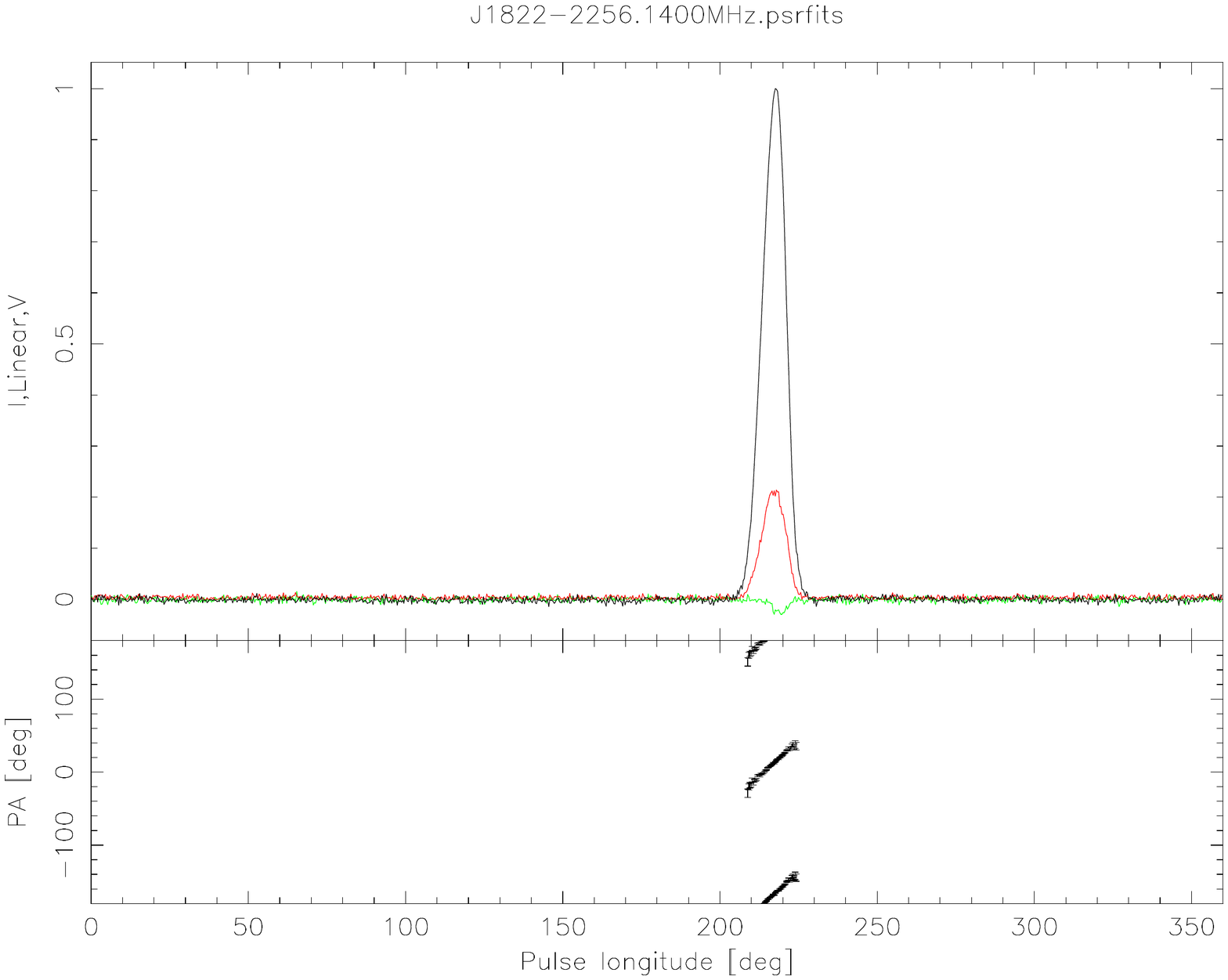}
	\includegraphics[width=\columnwidth, trim={1.5cm 0cm 4cm 2cm},clip]{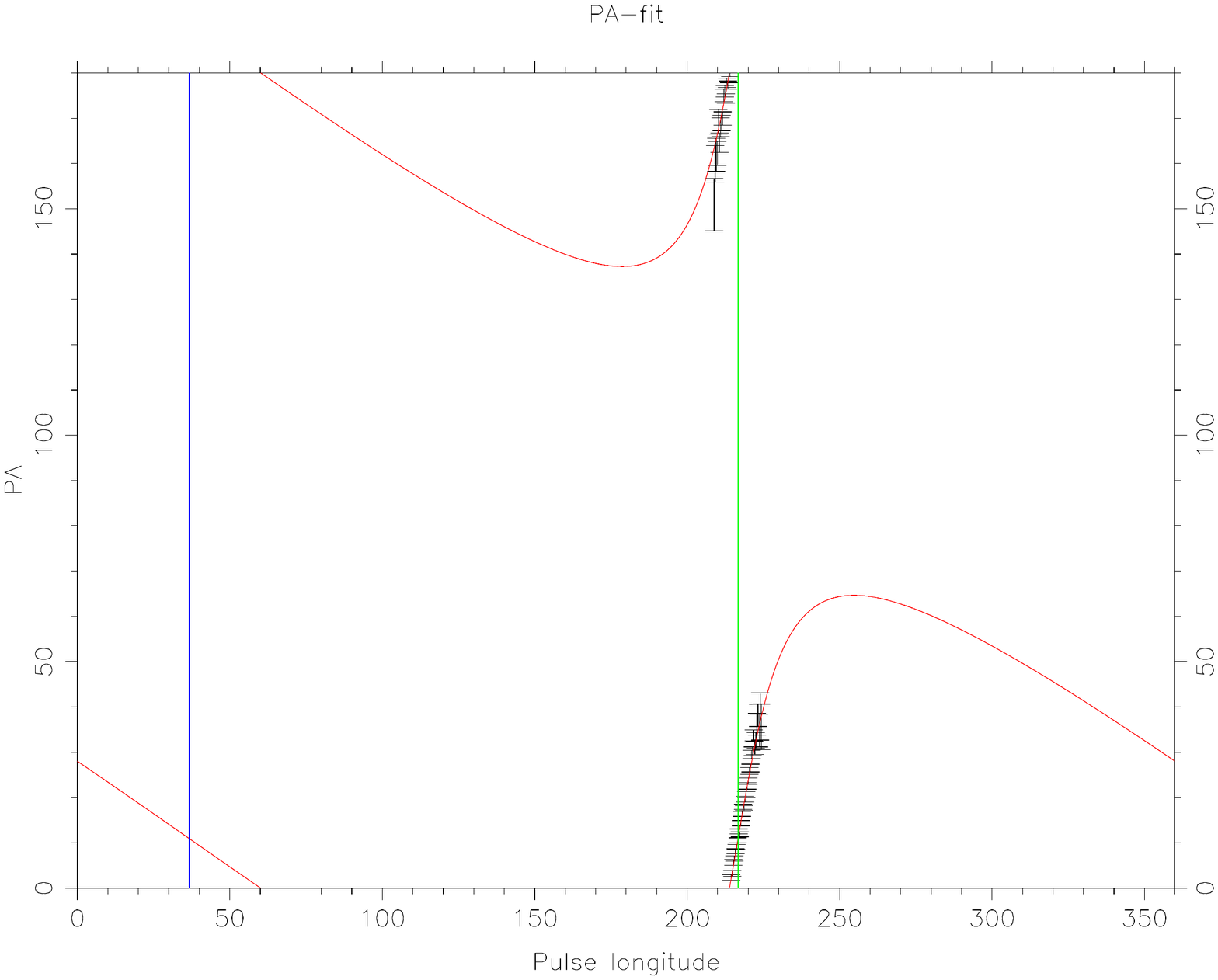}
    \caption{\textit{Top:} Average profile of J1822$-$2256 at 1400 MHz. \textit{Bottom:} Polarisation Angle fit for 1400 MHz profile of J1822$-$2256 obtained from Parkes data. The green line denotes the zero crossing point of PA, and the blue line is $180^{\circ}$ offset from that.}
    \label{fig:pafit}
\end{figure}

To constrain the pulsar geometry and determine the $\alpha$ and $\beta$ angles, we used the polarimetric data made available by \citet{10.1093/mnras/stx3095}. The data were taken using the Parkes radio telescope (recently given the indigenous name {\it Murriyang}) at 1.4 GHz with high time and frequency resolution. We used {\tt PSRSalsa} \citep{2018IAUS..337..424W} to fit the RVM and compute $\alpha$ and $\beta$ from the polarimetric data. For this pulsar, the value of $\alpha$ could not be well constrained. Fig. \ref{fig:pafit} shows the PA fitting for the  Parkes data with $\alpha = 23.9^{\circ}$ and $\beta = 5.7^{\circ}$, along with the average profile. 

Additionally, the steepest gradient (SG), which is defined as sin $\alpha$/sin $\beta$, can be used to estimate the profile shape. For a central cut of the emission beam, the profile will have multiple components with core and conal emission; consequently, the value of SG would be higher. For our case, the value of SG is $\sim 4 ^{\circ} / ^{\circ}$, implying a conal cut of the emission beam.

\subsection{Determination of emission heights for the emission modes} \label{sec:emheight}

Fig. \ref{fig:gfit} shows how the two components of the average profile change with an increase in frequency; as seen from this figure, the profile becomes narrower with an increase in frequency. This phenomenon, common to many pulsars, is usually ascribed to, the radius-to-frequency mapping (RFM, \citealp{1978ApJ...222.1006C}), which suggests that emissions at different frequencies originate at different altitudes in the pulsar magnetosphere. The association of higher frequency emission with lower altitudes (and the lower frequency with higher emission altitudes) comes from the dipolar shape of the magnetic field lines expected in the emission region, since tangents to the field lines subtend a smaller angle to the magnetic axis closer to the surface, giving rise to a smaller conal opening angle.

To calculate the height above the pulsar surface at which the emission at a particular frequency arises, we need to calculate the half beam opening angle, $\Gamma$, which is determined as the angle between the magnetic axis and the cone of emission. Since the cone has a certain thickness, one can consider the half opening angle as that corresponds either to the inner or outer edge of the cone or somewhere in the middle. Here we choose the outer edge since the corresponding points in the profile (i.e. the leading and trailing edges) are relatively easy to measure, even when the components start to merge at higher frequencies. In particular, we assume that the 10\% pulse width, $W_{10}$, is a good approximation to where the emission from the outer conal edge appears. Using the three sides of a spherical triangle formed between the magnetic axis, rotation axis, and line of sight, we can write

\begin{equation}
    \cos \Gamma = \cos \alpha \cos(\alpha + \beta) + \sin \alpha \sin(\alpha + \beta) \cos \left( \frac{W_{10}}{2} \right)
    \label{eqn:hba}
\end{equation}

We used $\alpha = 23.9^{\circ}$ and $\beta = 5.7^{\circ}$ for this case(as calculated using the 1.4 GHz Parkes data). Once the half beam opening angle has been obtained from eqn. \ref{eqn:hba}, we can calculate the angle subtended by the emission site with the magnetic axis, $\theta$, as \citep{Gangadhara2004} 

\begin{equation}
    \sin^2\theta = - \frac{1}{6} \left( \cos \Gamma \sqrt{8 + \cos^2 \Gamma} + \cos^2 \Gamma - 4 \right)
\end{equation}

\begin{figure}
	\includegraphics[width=\columnwidth]{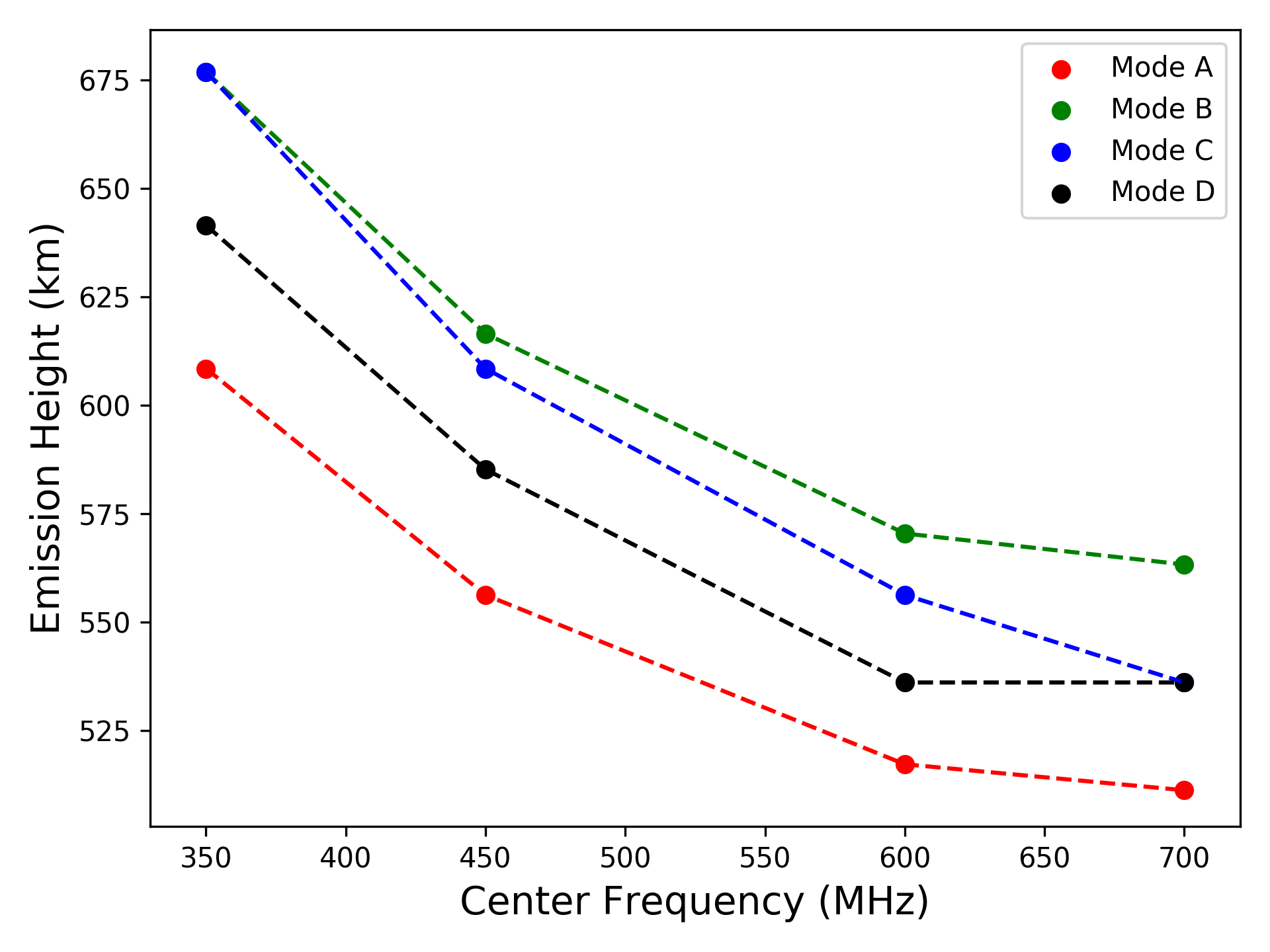}
    \caption{Plot of emission heights for different modes with frequency. Different colours represent the different subpulse drifting modes.}
    \label{fig:height}
\end{figure}

Assuming that the radius of the neutron star is much smaller than the light cylinder radius ($r_p \ll R_{LC}$), we can calculate the emission height, $r$, normalised to the light cylinder radius ($R_{LC}$) as
\begin{equation}
    \frac{r}{R_{LC}} \approx \frac{sin^2 \theta}{s^2}
\end{equation}
where $s$ is the footprint parameter, and the location of the footprint is on the neutron star surface, which is assumed to be spherical. We have assumed that the radiation from the outer edge of the cone originates from the last open field line, for which $s = 1$. Using the viewing geometry obtained from the Parkes data, and the light cylinder radius, $R_{lc} = \frac{c P_1}{2 \pi} \cong 89,450$ km, we calculated the emission heights for all modes separately at the four central frequencies. Fig. \ref{fig:height} shows the emission heights at multiple observation frequencies. Here, the different colours correspond to the four modes of subpulse drifting found in our study. We used the $\alpha$ and $\beta$ values obtained from polarimetric analysis of the high-quality Parkes data, which are different from the ones calculated by \citet{2018MNRAS.476.1345B}. The error bars could not be calculated for this case due to the poorly constrained pulsar geometry. However, the trend in emission heights across the different modes, throughout multiple trials of pulsar geometry, remained the same. The possible sources of uncertainty on the emission heights could be the error in constraining $\alpha$ and $\beta$ values and the error due to signal-to-noise of the average profile, propagating in the 10\% profile widths. 

%%%%%%%%%%%%%%%%%%%%%%%%%%%%%%%%%%%%%%%%%%%%%%%%%%%%%%%%%%%%%%%%%%%%%

%% file: discussion.tex
\section{Discussion}\label{sec:discussion}

In this paper, we have presented the results from the first simultaneous wide-band multi-frequency study on the pulsar J1822$-$2256 using the uGMRT. This pulsar is known to exhibit subpulse drifting, mode changing, as well as nulling and is, therefore, a promising candidate to study the pulsar emission mechanism. As discussed in the previous section, our analysis suggests the presence of new features in some of the drift modes of emission, which were not reported by previous work. We also present insightful results relating to mode dependent emission heights. In this section, we discuss the implications of our results on the pulsar emission process, in an effort to shed some light on the possible modifications needed in the carousel model for this pulsar. Specifically, we consider the spark carousel geometry and the relation between different modes and their inferred emission heights. 

To summarise, we have found four subpulse drifting modes for the pulsar J1822$-$2256, including some new features that were not studied before. Modes A and C were commonly found in all previously published literature, with our $P_3$ values consistent within their error bars. We found another subpulse drifting mode at $P_3$ $\sim 5.8\;P_1$ with interesting features. This mode could be similar to mode B, which was categorised as a non-drifting mode by \citet{2018MNRAS.476.1345B}. It was also referred to as mode C in \citet{2018IAUS..337..348J}. Results reported by \citet{2018MNRAS.476.1345B} were from observations made at 322 MHz with 33 MHz bandwidth. However, having higher sensitivity due to factor of six increase in the observing bandwidth, we and \citet{2018IAUS..337..348J} have found the drifting nature in mode B (mode C for \citet{2018IAUS..337..348J}). \citet{2009A&A...506..865S} also reported the presence of a 'nulling or rapidly changing drift mode', which could be same as the mode B in our case. In addition, we have found another subpulse drifting mode - mode D. Even though \citet{2018MNRAS.476.1345B} found a transitional mode A at the same $P_3$ as our mode D, the primary characteristic between the two modes is different. In the case of \citet{2018MNRAS.476.1345B}, they claimed that this mode occurred after mode A in most cases, whereas, in our data, most of the occurrences of mode D are preceded by a null. With such varying values for $P_3$ for all the modes and order of appearance of modes, the pulsar possibly shows an evolving subpulse drifting behaviour altogether. It is also possible that observations made at different epochs find the magnetosphere in different dynamical states resulting in observed differences in subpulse drift properties by the individual studies.

\subsection{Spark calculation}\label{sec:spark}

We now attempt to determine the number of sparks in the carousel by assuming they are equally spaced in the magnetic azimuthal direction. The behaviour of drift bands is determined by the number of sparks, $n$, and the rotation rate of the carousel, $P_4$, i.e. the time for the sparks to complete one revolution around the polar cap. However, the observed drifting could be aliased due to an under-sampling of the subpulse motion or the observer's inability to distinguish between subpulses \citep{2003A&A...399..223V}. Because we can only observe the position of a subpulse once per pulsar rotation, it is rather difficult to determine the real speed of carousel rotation exactly. 

We have found different $P_3$ values for the different modes of J1822$-$2256, which can be explained by a change in the number of sparks or the carousel rotation rate. In the general case, the carousel rotation rate may be aliased with the star's rotation, with the observed $P_3$ obeying the relation

\begin{equation}
    \frac{1}{\overline{P_3}} =  \biggm \lvert \frac{n}{\overline{P_4}} - k \biggm \lvert
    \label{eqn:p4}
\end{equation}

\noindent
Here we have followed the overline notation used by \citet{2019ApJ...883...28M}, where $\overline{P_3}$ is $P_3/P_1$ and $\overline{P_4}$ is $P_4/P_1$. The parameter $k = [n/\overline{P_4}]$ is the aliasing order, with square brackets denoting rounding off to the nearest integer. 

Because of the free choice of $P_4$ for each mode, it is not easy to find the correct aliasing order, $k$. However, the carousel rotation speed cannot easily change its magnitude or direction on the timescales in which the drift modes change, which, for our case, is around one pulsar rotation. For PSR B1918+19, \citet{2013MNRAS.433..445R} suggested keeping $P_4$ constant while allowing a change in $n$. In that case, $k$ cannot be 0 for all drift modes since otherwise, the drift rate would be the same for all four modes.

The challenge is therefore to find values for $\overline{P_4}$ for which integer values of $n_A$, $n_B$, etc. can be found to predict the correct values of $\overline{P_3}$ from eqn.~\eqref{eqn:p4}. Given the ranges of $\overline{P_3}$ values available for each mode (as per Fig. \ref{fig:P3full}), it is possible that many solutions can be found that are broadly consistent with the $\overline{P_3}$ values for each mode. Given this, we will opt to favour $\overline{P_4}$ values that are closer to the theoretical value predicted by the Ruderman-Sutherland model, which is
\begin{equation}
    \overline{P_4}_{,RS} \approx 5.7 \times \left( \frac{P_1}{s} \right)^{-3/2} \left( \frac{\Dot{P}}{10^{-15}} \right)^{1/2}
    \label{eqn:P4RS}
\end{equation}

where ${\dot P}$ is the spin-down rate (i.e. the first derivative if the spin period), which is $1.35439 \times  10^{-15} \, s\,s^{-1}$ for PSR J1822$-$2256 \citealp{2005yCat.7245....0M}). For J1822$-$2256, eqn.~\eqref{eqn:P4RS} gives $\overline{P_4}_{,RS} \approx 2.79$.

We searched for solutions by brute force, i.e., testing every value $0 < \overline{P_4} \lesssim 30$ at increments of $0.01$. For each $\overline{P_4}$, we used eqn. ~\eqref{eqn:p4} to find the best values of $n$ that produce the nominal $\overline{P_3}$ values for each mode. We repeated this exercise for each $k$ in $[-2, -1, 1, 2]$, i.e. up to second-order aliasing.

Interestingly, we were not able to find any solution that was consistent with all four modes by this method. One possible reason for this is our implicit requirement that all four modes belong to the same aliasing order may not hold; it may well be the case that some modes have (for example) $k=1$, while others have $k=2$. Nevertheless, we found that solutions with mixed aliasing orders usually required the number of sparks to jump by a large amount between the modes, which seems counter to the types of solutions obtained for other pulsars, for which the number of sparks only ever jumps by one or two during mode switches \citep[e.g.][and references therein]{2019ApJ...883...28M}.

When we restricted our search to just those that satisfied three of the four modes, we were able to find solutions for both first and second-order solutions. The solutions with the smallest $\overline{P_4}$ values tended to be in the range $13 \lesssim \overline{P_4} \lesssim 16$. Table \ref{tab:p4} shows a few representative solutions that were found for modes A, B, and C, for both first and second order aliasing.
\begin{table}
	\centering
	\caption{Carousel parameters for J1822$-$2256}
	\label{tab:p4}
	\begin{tabular}{ccc|ccc|cc} 
		\hline
		$n_A$ & $n_B$ & $n_C$ & $\overline{P_{3A}}$ & $\overline{P_{3B}}$ & $\overline{P_{3C}}$ & $k$ & $\sim \overline{P_4}$ \\
		\hline
		14 & 12 & 13 & 17.7 & 5.2 & 8.1 & +1 & 14.84 \\
        31 & 33 & 32 & 18.0 & 5.3 & 8.2 & +2 & 15.08 \\
		\hline
	\end{tabular}
\end{table}

It is interesting to note that, for these solutions, $\overline{P_4}$ is very close to the observed $\overline{P_{3D}} \approx 14$, a scenario in which the $\overline{P_3}$ value predicted by eqn.~\eqref{eqn:p4} becomes very sensitive to the precise value of $\overline{P_4}$. A situation like this may go some way to explaining the difficulty of finding solutions that fit all four modes simultaneously. However, in the case of PSR J1822$-$2256, our ability to identify the correct solution is more likely to be limited by the uncertainties on the $\overline{P_3}$ values, rather than the precision used in the search procedure. Nevertheless, we reiterate that, among the solutions found by the above procedure, the most common values found for $\overline{P_4}$ fell in the range $13 \lesssim \overline{P_4} \lesssim 16$.

\subsection{Modes and implications for pulsar emission}

In this study, we have identified four distinct subpulse drifting emission modes in our observations and calculated the possible spark configuration for three out of four drift modes. With these calculated number of sparks, occurrence fraction of each mode, and the average burst lengths, we can comment on the steadiness of modes.

With a $P_3$ of $\sim 18\;P_1$, mode A has the largest occurrence fraction and average burst length. Therefore, sparks giving rise to mode A seem to be comparatively long-lasting. The two peaks in the average profile, corresponding to the leading and the trailing edge of the profile, are closest for mode A. 

The bursts of mode B, despite being short, show partially organised drifting, along with the largest variation in $P_3$ values (Fig. \ref{fig:P3full}). This mode is possibly the same as the non-drifting mode mentioned in \citet{2018MNRAS.476.1345B} and mode C for \citet{2018IAUS..337..348J}. According to our classification, the defining feature for this mode was the apparent lack of emission in some parts of the drift sequence (Fig. \ref{fig:modeps}). The average profile of this mode shows two widely separated components, with a generally large $P_2$ value and the broken drift sequences. According to our calculation, mode B has the smallest burst length, and it shows the second-highest occurrence fraction. Using these characteristics, we can infer that the sparks responsible for mode B are the most unstable, i.e., short-lived. This mode was also seen to be the most susceptible to pseudo nulls, which are generally two to three pulses wide. For this pulsar, most of the pseudo nulls are seen around mode B. This could be due to relatively low stability of sparks; therefore, higher chances of the observer's line of sight passing through a minimum between sub-beams.

Mode C in our observations occasionally shows the occurrence of drifting that is akin to mode B, i.e. broken drift sequences. However, the primary difference between the two modes is that the $P_3$ value is larger for mode C, and the drift sequences do not generally have more than one break. Compared to mode C mentioned in the previously published literature, this mode shows a slightly different $P_3$ value but still within the error bars. On average, in our data, the burst length for mode C is 15 pulses, which is considerably higher than that of mode B. Frequent transitions between mode B and mode C were also noted and are generally bridged by the occurrence of a pseudo-null. These pseudo nulls, as mentioned in \citet{2019ApJ...870..110M} for PSR B1237+25, may represent the chance positioning of the observer's line of sight across the minima between the sub-beams in the carousel, as against the physical cessation of the emission in the actual nulls.

The least observed mode in our data is mode D, with an occurrence fraction of about 6\%. \citet{2018MNRAS.476.1345B} found a transitional-A mode at a similar $P_3$, which was seen occasionally when the pulsar transitions from mode A to some other mode. Interestingly, for our case, most of the occurrences of this mode were preceded by a null. We did not find any correlation between the occurrence of mode D and mode A. The drifting pattern of this mode looks similar to mode A, but with a different $P_3$ value of $14\;P_1$. Given the low occurrence rate and a large average burst length, we can interpret that the sparks responsible for this mode are long-lasting but stay dormant for longer times and activate less frequently.

The phenomenon of subpulse drifting is thought to be due to the rotation of conal sparks around the magnetic axis. When the line of sight cuts through the conal part of the emission beam, subpulse drifting is observed. The closer the line of sight moves towards the magnetic pole, the less prominent will be the observers drifting. In our case, drifting is observed throughout the full profile, implying that the observer's line of sight is possibly cutting through the conal sparks. This is also in accordance with the inference made by \citet{2018MNRAS.476.1345B}, where they have used the steepest gradient point of the polarisation position angle to suggest that the observer's line of sight cuts the emission beam tangentially. If the line of sight cuts through the core region of the pulsar beam, closer to the magnetic pole, no drifting will be observed in the section corresponding to the core region. This implies that the mean position of the profile component will not change. However, as shown in Fig. \ref{fig:gfit}, both the profile components, trailing and leading, are drifting with a certain trend in frequency, implying that the observer's line of sight is cutting through the conal sparks. 

In addition, our analysis reveals a different evolution of the leading and trailing profile components with frequency. As shown in Fig. \ref{fig:gfit}, the leading components show a higher degree of evolution than the trailing component. We also calculated the aberration and retardation effects which could possibly affect the frequency-dependent shifts. However, our calculations (based on the heights inferred in the following section) showed the effects to be too small to cause any noticeable change. Therefore, such behaviour can be explained with a carousel that might be offset from the magnetic axis or has an irregular shape \citep{2013MNRAS.431.2756J}. In such a scenario, the sparks in the trailing component could be closer to the magnetic axis and therefore show a lower degree of frequency evolution than the sparks in the leading profile component, which might be further away from the magnetic pole.

\subsection{Emission heights for modes}

Using the radius to frequency mapping explanation, we extracted the information about the emission heights from our multi-frequency observations of J1822$-$2256. Based on our calculations, we find that this pulsar's emission heights depend upon the pulsar geometry, specifically $\alpha$, which is not well constrained. Using the Parkes 1400 MHz data, we have constrained the emission heights to $\sim$500 to 700 km in the observed frequency range with the geometry that we find most plausible.

As evident from the calculation in section \ref{sec:emheight}, the emission heights are related to the profile width. Consequently, one can notice that the profile for mode A is narrower, which can be because the open magnetic field lines are not as much separated at lower heights. The same argument follows for the other modes as well, where mode B shows a broader profile than mode C and is emitted from a higher emission height.

Interestingly, we also noticed that the emission heights are anti-correlated with $P_3$. Thus, mode B, which has the lowest $P_3$ value, is emitted from a higher altitude in the pulsar magnetosphere than mode A, which has a higher $P_3$ value and is inferred to have lower emission heights. A similar trend of heights is followed at all the frequencies as well as multiple trial geometries. Considering only the emission from modes A, B, and C, (accounting for almost 94\% of the total pulsar emission), some spark solutions infer that the emission heights are proportional to the number of sparks calculated for each of the modes. However, our spark analysis shows multiple possible solutions for this pulsar. Hence, other clues, such as the modal profile width, $P_2$, $P_3$, etc., can help determine which spark solution is more likely. To the best of our knowledge it is for the first time that such a trend, between $P_3$ and emission height has been seen in a sub-pulse drifting pulsar. It can potentially serve as a vital clue for understanding the pulsar emission process and the spark behaviour in the pulsar magnetosphere.

%% file: conclusion.tex
\section{Summary} \label{sec:conclusion}

In this paper, we have presented the results from the first wide-bandwidth single pulse study of PSR J1822$-$2256, covering a frequency range of 300-750 MHz. Observations were made using the upgraded GMRT in the dual-frequency mode. Our analysis confirms the previously reported subpulse drifting modes and uncovered some additional features which were not studied earlier, possibly due to the limited sensitivity of those observations. Using our observations, we have been able to study the drifting behaviour of this pulsar over a wider frequency range. We have presented a new subpulse drifting classification for this pulsar, based on the modes' average profiles and the corresponding $P_3$ values. We identify at least four different subpulse drifting modes with clearly distinguishable characteristics. 

Our analysis also reveals that the pulsar exhibits short-duration pseudo nulls in addition to genuine nulls. On an individual pulse level, both types of nulls show similar noise-like behaviour. However, a low-level emission is observed in the average profile constructed from pseudo nulls across the observing frequency range. These pseudo nulls are primarily present in the modes B and C, which have the lowest $P_3$ values and occasionally exhibit the apparent lack of emission in the drift bands. Previous studies suggested the presence of such pseudo nulls due to the observer's line of sight passing through a minimum of sparks.
    
We have also studied the frequency-dependent subpulse behaviour of different modes by segmenting the 200 MHz observing bandwidth into 100 MHz sub-bands. This exercise resulted in sub-banded data points to investigate the frequency evolution of the pulsar. Based on an observed evolution of profile components, where the leading components seem to evolve faster than the trailing component, we have proposed that the underlying carousel could either be offset from the magnetic pole or has an irregular shape. Such a carousel geometry would imply certain sparks closer to the magnetic pole, showing a lower degree of frequency evolution.

With the resultant multi-frequency measurements, we were also able to demonstrate the confirmation of radius-to-frequency mapping for each of the modes across the frequency range of our data. Further, we used the modal frequency evolution to estimate the emission heights. Our results show that the emission heights at different frequencies are dependent upon the mode. Our analysis has revealed an anti-correlation between the inferred emission heights and the measured $P_3$ values. Interestingly, solutions were found in which the inferred emission heights were directly correlated to the number of sparks. If such a correlation is true, this will imply that the numbers of sparks in a mode are directly related to the emission height, for at least 94 \% of the emission for this pulsar. Our results may provide vital clues toward understanding the pulsar emission mechanism and refining the spark model.

%% file: main.bbl
\begin{thebibliography}{}
\makeatletter
\relax
\def\mn@urlcharsother{\let\do\@makeother \do\$\do\&\do\#\do\^\do\_\do\%\do\~}
\def\mn@doi{\begingroup\mn@urlcharsother \@ifnextchar [ {\mn@doi@}
  {\mn@doi@[]}}
\def\mn@doi@[#1]#2{\def\@tempa{#1}\ifx\@tempa\@empty \href
  {http://dx.doi.org/#2} {doi:#2}\else \href {http://dx.doi.org/#2} {#1}\fi
  \endgroup}
\def\mn@eprint#1#2{\mn@eprint@#1:#2::\@nil}
\def\mn@eprint@arXiv#1{\href {http://arxiv.org/abs/#1} {{\tt arXiv:#1}}}
\def\mn@eprint@dblp#1{\href {http://dblp.uni-trier.de/rec/bibtex/#1.xml}
  {dblp:#1}}
\def\mn@eprint@#1:#2:#3:#4\@nil{\def\@tempa {#1}\def\@tempb {#2}\def\@tempc
  {#3}\ifx \@tempc \@empty \let \@tempc \@tempb \let \@tempb \@tempa \fi \ifx
  \@tempb \@empty \def\@tempb {arXiv}\fi \@ifundefined
  {mn@eprint@\@tempb}{\@tempb:\@tempc}{\expandafter \expandafter \csname
  mn@eprint@\@tempb\endcsname \expandafter{\@tempc}}}

\bibitem[\protect\citeauthoryear{{Backer}}{{Backer}}{1970}]{1970Natur.228...42B}
{Backer} D.~C.,  1970, \mn@doi [\nat] {10.1038/228042a0}, \href
  {https://ui.adsabs.harvard.edu/abs/1970Natur.228...42B} {228, 42}

\bibitem[\protect\citeauthoryear{{Backer}}{{Backer}}{1973}]{1973ApJ...182..245B}
{Backer} D.~C.,  1973, \mn@doi [\apj] {10.1086/152134}, \href
  {https://ui.adsabs.harvard.edu/abs/1973ApJ...182..245B} {182, 245}

\bibitem[\protect\citeauthoryear{{Basu} \& {Mitra}}{{Basu} \&
  {Mitra}}{2018}]{2018MNRAS.476.1345B}
{Basu} R.,  {Mitra} D.,  2018, \mn@doi [\mnras] {10.1093/mnras/sty297}, \href
  {https://ui.adsabs.harvard.edu/abs/2018MNRAS.476.1345B} {476, 1345}

\bibitem[\protect\citeauthoryear{{Cordes}}{{Cordes}}{1978}]{1978ApJ...222.1006C}
{Cordes} J.~M.,  1978, \mn@doi [\apj] {10.1086/156218}, \href
  {https://ui.adsabs.harvard.edu/abs/1978ApJ...222.1006C} {222, 1006}

\bibitem[\protect\citeauthoryear{{Drake} \& {Craft}}{{Drake} \&
  {Craft}}{1968}]{1968Natur.220..231D}
{Drake} F.~D.,  {Craft} H.~D.,  1968, \mn@doi [\nat] {10.1038/220231a0}, \href
  {https://ui.adsabs.harvard.edu/abs/1968Natur.220..231D} {220, 231}

\bibitem[\protect\citeauthoryear{{Edwards} \& {Stappers}}{{Edwards} \&
  {Stappers}}{2003}]{2003A&A...410..961E}
{Edwards} R.~T.,  {Stappers} B.~W.,  2003, \mn@doi [\aap]
  {10.1051/0004-6361:20031326}, \href
  {https://ui.adsabs.harvard.edu/abs/2003A&A...410..961E} {410, 961}

\bibitem[\protect\citeauthoryear{Gajjar, Joshi  \& Kramer}{Gajjar
  et~al.}{2012}]{10.1111/j.1365-2966.2012.21296.x}
Gajjar V.,  Joshi B.~C.,   Kramer M.,  2012, \mn@doi [Monthly Notices of the
  Royal Astronomical Society] {10.1111/j.1365-2966.2012.21296.x}, 424, 1197

\bibitem[\protect\citeauthoryear{Gangadhara}{Gangadhara}{2004}]{Gangadhara2004}
Gangadhara R.~T.,  2004, \mn@doi [Astrophysical Journal] {10.1086/420961}, 609,
  335

\bibitem[\protect\citeauthoryear{{Gold}}{{Gold}}{1969}]{1969Natur.221...25G}
{Gold} T.,  1969, \mn@doi [\nat] {10.1038/221025a0}, \href
  {https://ui.adsabs.harvard.edu/abs/1969Natur.221...25G} {221, 25}

\bibitem[\protect\citeauthoryear{{Gupta} et~al.,}{{Gupta}
  et~al.}{2017}]{2017CSci..113..707G}
{Gupta} Y.,  et~al., 2017, Current Science, \href
  {https://ui.adsabs.harvard.edu/abs/2017CSci..113..707G} {113, 707}

\bibitem[\protect\citeauthoryear{{Hewish}, {Bell}, {Pilkington}, {Scott}  \&
  {Collins}}{{Hewish} et~al.}{1968}]{1968Natur.217..709H}
{Hewish} A.,  {Bell} S.~J.,  {Pilkington} J.~D.~H.,  {Scott} P.~F.,   {Collins}
  R.~A.,  1968, \mn@doi [\nat] {10.1038/217709a0}, \href
  {https://ui.adsabs.harvard.edu/abs/1968Natur.217..709H} {217, 709}

\bibitem[\protect\citeauthoryear{{Hotan}, {van Straten}  \&
  {Manchester}}{{Hotan} et~al.}{2004}]{2004PASA...21..302H}
{Hotan} A.~W.,  {van Straten} W.,   {Manchester} R.~N.,  2004, \mn@doi [\pasa]
  {10.1071/AS04022}, \href
  {https://ui.adsabs.harvard.edu/abs/2004PASA...21..302H} {21, 302}

\bibitem[\protect\citeauthoryear{{Huguenin}, {Taylor}  \& {Troland}}{{Huguenin}
  et~al.}{1970}]{1970ApJ...162..727H}
{Huguenin} G.~R.,  {Taylor} J.~H.,   {Troland} T.~H.,  1970, \mn@doi [\apj]
  {10.1086/150704}, \href
  {https://ui.adsabs.harvard.edu/abs/1970ApJ...162..727H} {162, 727}

\bibitem[\protect\citeauthoryear{Johnston \& Kerr}{Johnston \&
  Kerr}{2017}]{10.1093/mnras/stx3095}
Johnston S.,  Kerr M.,  2017, \mn@doi [Monthly Notices of the Royal
  Astronomical Society] {10.1093/mnras/stx3095}, 474, 4629

\bibitem[\protect\citeauthoryear{{Jones}}{{Jones}}{2013}]{2013MNRAS.431.2756J}
{Jones} P.~B.,  2013, \mn@doi [\mnras] {10.1093/mnras/stt372}, \href
  {https://ui.adsabs.harvard.edu/abs/2013MNRAS.431.2756J} {431, 2756}

\bibitem[\protect\citeauthoryear{{Joshi}, {Naidu}, {Gajjar}  \&
  {Wright}}{{Joshi} et~al.}{2018}]{2018IAUS..337..348J}
{Joshi} B.~C.,  {Naidu} A.,  {Gajjar} V.,   {Wright} G. A.~E.,  2018, in
  {Weltevrede} P.,  {Perera} B.~B.~P.,  {Preston} L.~L.,   {Sanidas} S.,  eds,
  ~ Vol. 337, Pulsar Astrophysics the Next Fifty Years. pp 348--349,
  \mn@doi{10.1017/S1743921317008390}

\bibitem[\protect\citeauthoryear{{Maan}}{{Maan}}{2019}]{2019ApJ...870..110M}
{Maan} Y.,  2019, \mn@doi [\apj] {10.3847/1538-4357/aaf41c}, \href
  {https://ui.adsabs.harvard.edu/abs/2019ApJ...870..110M} {870, 110}

\bibitem[\protect\citeauthoryear{{Manchester}, {Hobbs}, {Teoh}  \&
  {Hobbs}}{{Manchester} et~al.}{2005}]{2005yCat.7245....0M}
{Manchester} R.~N.,  {Hobbs} G.~B.,  {Teoh} A.,   {Hobbs} M.,  2005, VizieR
  Online Data Catalog, \href
  {https://ui.adsabs.harvard.edu/abs/2005yCat.7245....0M} {p. VII/245}

\bibitem[\protect\citeauthoryear{{McSweeney}, {Bhat}, {Tremblay}, {Deshpande}
  \& {Ord}}{{McSweeney} et~al.}{2017}]{2017ApJ...836..224M}
{McSweeney} S.~J.,  {Bhat} N.~D.~R.,  {Tremblay} S.~E.,  {Deshpande} A.~A.,
  {Ord} S.~M.,  2017, \mn@doi [\apj] {10.3847/1538-4357/aa5c35}, \href
  {https://ui.adsabs.harvard.edu/abs/2017ApJ...836..224M} {836, 224}

\bibitem[\protect\citeauthoryear{{McSweeney}, {Bhat}, {Wright}, {Tremblay}  \&
  {Kudale}}{{McSweeney} et~al.}{2019}]{2019ApJ...883...28M}
{McSweeney} S.~J.,  {Bhat} N.~D.~R.,  {Wright} G.,  {Tremblay} S.~E.,
  {Kudale} S.,  2019, \mn@doi [\apj] {10.3847/1538-4357/ab3a97}, \href
  {https://ui.adsabs.harvard.edu/abs/2019ApJ...883...28M} {883, 28}

\bibitem[\protect\citeauthoryear{{Melrose}}{{Melrose}}{2017}]{2017RvMPP...1....5M}
{Melrose} D.~B.,  2017, \mn@doi [Reviews of Modern Plasma Physics]
  {10.1007/s41614-017-0007-0}, \href
  {https://ui.adsabs.harvard.edu/abs/2017RvMPP...1....5M} {1, 5}

\bibitem[\protect\citeauthoryear{{Naidu}, {Joshi}, {Manoharan}  \&
  {KrishnaKumar}}{{Naidu} et~al.}{2017}]{2017A&A...604A..45N}
{Naidu} A.,  {Joshi} B.~C.,  {Manoharan} P.~K.,   {KrishnaKumar} M.~A.,  2017,
  \mn@doi [\aap] {10.1051/0004-6361/201629937}, \href
  {https://ui.adsabs.harvard.edu/abs/2017A&A...604A..45N} {604, A45}

\bibitem[\protect\citeauthoryear{{Qiao}, {Lee}, {Zhang}, {Xu}  \&
  {Wang}}{{Qiao} et~al.}{2004}]{2004ApJ...616L.127Q}
{Qiao} G.~J.,  {Lee} K.~J.,  {Zhang} B.,  {Xu} R.~X.,   {Wang} H.~G.,  2004,
  \mn@doi [\apjl] {10.1086/426862}, \href
  {https://ui.adsabs.harvard.edu/abs/2004ApJ...616L.127Q} {616, L127}

\bibitem[\protect\citeauthoryear{{Radhakrishnan} \& {Cooke}}{{Radhakrishnan} \&
  {Cooke}}{1969}]{1969ApL.....3..225R}
{Radhakrishnan} V.,  {Cooke} D.~J.,  1969, \aplett, \href
  {https://ui.adsabs.harvard.edu/abs/1969ApL.....3..225R} {3, 225}

\bibitem[\protect\citeauthoryear{{Rankin}}{{Rankin}}{1986}]{1986ApJ...301..901R}
{Rankin} J.~M.,  1986, \mn@doi [\apj] {10.1086/163955}, \href
  {https://ui.adsabs.harvard.edu/abs/1986ApJ...301..901R} {301, 901}

\bibitem[\protect\citeauthoryear{{Rankin}, {Wright}  \& {Brown}}{{Rankin}
  et~al.}{2013}]{2013MNRAS.433..445R}
{Rankin} J.~M.,  {Wright} G. A.~E.,   {Brown} A.~M.,  2013, \mn@doi [\mnras]
  {10.1093/mnras/stt739}, \href
  {https://ui.adsabs.harvard.edu/abs/2013MNRAS.433..445R} {433, 445}

\bibitem[\protect\citeauthoryear{{Reddy} et~al.,}{{Reddy}
  et~al.}{2017}]{2017JAI.....641011R}
{Reddy} S.~H.,  et~al., 2017, \mn@doi [Journal of Astronomical Instrumentation]
  {10.1142/S2251171716410117}, \href
  {https://ui.adsabs.harvard.edu/abs/2017JAI.....641011R} {6, 1641011}

\bibitem[\protect\citeauthoryear{{Redman}, {Wright}  \& {Rankin}}{{Redman}
  et~al.}{2005}]{2005MNRAS.357..859R}
{Redman} S.~L.,  {Wright} G. A.~E.,   {Rankin} J.~M.,  2005, \mn@doi [\mnras]
  {10.1111/j.1365-2966.2005.08672.x}, \href
  {https://ui.adsabs.harvard.edu/abs/2005MNRAS.357..859R} {357, 859}

\bibitem[\protect\citeauthoryear{Ruderman}{Ruderman}{1972}]{doi:10.1146/annurev.aa.10.090172.002235}
Ruderman M.,  1972, \mn@doi [Annual Review of Astronomy and Astrophysics]
  {10.1146/annurev.aa.10.090172.002235}, 10, 427

\bibitem[\protect\citeauthoryear{{Ruderman} \& {Sutherland}}{{Ruderman} \&
  {Sutherland}}{1975}]{1975ApJ...196...51R}
{Ruderman} M.~A.,  {Sutherland} P.~G.,  1975, \mn@doi [\apj] {10.1086/153393},
  \href {https://ui.adsabs.harvard.edu/abs/1975ApJ...196...51R} {196, 51}

\bibitem[\protect\citeauthoryear{{Serylak}, {Stappers}  \&
  {Weltevrede}}{{Serylak} et~al.}{2009}]{2009A&A...506..865S}
{Serylak} M.,  {Stappers} B.~W.,   {Weltevrede} P.,  2009, \mn@doi [\aap]
  {10.1051/0004-6361/200912453}, \href
  {https://ui.adsabs.harvard.edu/abs/2009A&A...506..865S} {506, 865}

\bibitem[\protect\citeauthoryear{{Smits}, {Mitra}  \& {Kuijpers}}{{Smits}
  et~al.}{2005}]{2005A&A...440..683S}
{Smits} J.~M.,  {Mitra} D.,   {Kuijpers} J.,  2005, \mn@doi [\aap]
  {10.1051/0004-6361:20041626}, \href
  {https://ui.adsabs.harvard.edu/abs/2005A&A...440..683S} {440, 683}

\bibitem[\protect\citeauthoryear{{Swarup}, {Ananthakrishnan}, {Kapahi}, {Rao},
  {Subrahmanya}  \& {Kulkarni}}{{Swarup} et~al.}{1991}]{1991CuSc...60...95S}
{Swarup} G.,  {Ananthakrishnan} S.,  {Kapahi} V.~K.,  {Rao} A.~P.,
  {Subrahmanya} C.~R.,   {Kulkarni} V.~K.,  1991, Current Science, \href
  {https://ui.adsabs.harvard.edu/abs/1991CuSc...60...95S} {60, 95}

\bibitem[\protect\citeauthoryear{{Taylor} \& {Huguenin}}{{Taylor} \&
  {Huguenin}}{1971}]{1971ApJ...167..273T}
{Taylor} J.~H.,  {Huguenin} G.~R.,  1971, \mn@doi [\apj] {10.1086/151030},
  \href {https://ui.adsabs.harvard.edu/abs/1971ApJ...167..273T} {167, 273}

\bibitem[\protect\citeauthoryear{Wang, Manchester  \& Johnston}{Wang
  et~al.}{2007}]{10.1111/j.1365-2966.2007.11703.x}
Wang N.,  Manchester R.~N.,   Johnston S.,  2007, \mn@doi [Monthly Notices of
  the Royal Astronomical Society] {10.1111/j.1365-2966.2007.11703.x}, 377, 1383

\bibitem[\protect\citeauthoryear{{Weltevrede}}{{Weltevrede}}{2018}]{2018IAUS..337..424W}
{Weltevrede} P.,  2018, in {Weltevrede} P.,  {Perera} B.~B.~P.,  {Preston}
  L.~L.,   {Sanidas} S.,  eds, ~ Vol. 337, Pulsar Astrophysics the Next Fifty
  Years. pp 424--425, \mn@doi{10.1017/S1743921317007311}

\bibitem[\protect\citeauthoryear{{Weltevrede}, {Edwards}  \&
  {Stappers}}{{Weltevrede} et~al.}{2006}]{2006A&A...445..243W}
{Weltevrede} P.,  {Edwards} R.~T.,   {Stappers} B.~W.,  2006, \mn@doi [\aap]
  {10.1051/0004-6361:20053088}, \href
  {https://ui.adsabs.harvard.edu/abs/2006A&A...445..243W} {445, 243}

\bibitem[\protect\citeauthoryear{{Weltevrede}, {Stappers}  \&
  {Edwards}}{{Weltevrede} et~al.}{2007}]{2007A&A...469..607W}
{Weltevrede} P.,  {Stappers} B.~W.,   {Edwards} R.~T.,  2007, \mn@doi [\aap]
  {10.1051/0004-6361:20066855}, \href
  {https://ui.adsabs.harvard.edu/abs/2007A&A...469..607W} {469, 607}

\bibitem[\protect\citeauthoryear{{van Leeuwen}, {Stappers}, {Ramachandran}  \&
  {Rankin}}{{van Leeuwen} et~al.}{2003}]{2003A&A...399..223V}
{van Leeuwen} A.~G.~J.,  {Stappers} B.~W.,  {Ramachandran} R.,   {Rankin}
  J.~M.,  2003, \mn@doi [\aap] {10.1051/0004-6361:20021630}, \href
  {https://ui.adsabs.harvard.edu/abs/2003A&A...399..223V} {399, 223}

\bibitem[\protect\citeauthoryear{{van Straten} \& {Bailes}}{{van Straten} \&
  {Bailes}}{2011}]{2011PASA...28....1V}
{van Straten} W.,  {Bailes} M.,  2011, \mn@doi [\pasa] {10.1071/AS10021}, \href
  {https://ui.adsabs.harvard.edu/abs/2011PASA...28....1V} {28, 1}

\makeatother
\end{thebibliography}
